\documentclass[a4paper,11pt]{article}

\usepackage{geometry}									
\usepackage[english]{babel}		
\usepackage[utf8]{inputenc}		
\usepackage[T1]{fontenc}		

\usepackage{lmodern}			

\usepackage{graphicx}
\usepackage{tabularx}
\usepackage{amsmath}
\usepackage{amssymb}
\usepackage{amsfonts}
\usepackage{multirow}
\usepackage{booktabs}
\usepackage{colortbl}
\usepackage{xcolor}
\graphicspath{{figures}}

\makeatletter
\let\oldabs\abs
\def\abs{\@ifstar{\oldabs}{\oldabs*}}
\let\oldnorm\norm
\def\norm{\@ifstar{\oldnorm}{\oldnorm*}}
\makeatother

\newcommand{\ve}[1]{\boldsymbol{#1}} 
\newcommand{\cm}{\mathcal{M}} 
\newcommand{\ct}{\mathcal{T}}

\newcommand{\vx}{\ve{x}}
\newcommand{\vy}{\ve{y}}
\newcommand{\vX}{\ve{X}}

\newcommand{\vXi}{\ve{\Xi}}
\newcommand{\vxi}{\ve{\xi}}

\usepackage{authblk}						

\title{Time-variant reliability using time-dependent surrogate models}

\author{Stefano Marelli}
\author{Styfen Sch\"ar}
\author{Bruno Sudret}
\affil{Chair of Risk, Safety and Uncertainty Quantification, ETH Z\"{u}rich, Switzerland}

\date{May 11, 2026}

\usepackage{microtype}			

\usepackage{indentfirst}		

\usepackage{caption}
\usepackage{subcaption}

\usepackage{natbib}
\usepackage{hyperref}
\hypersetup{
pdftitle = {Time-variant reliability using time-dependent surrogate models},
pdfauthor = {Stefano Marelli, Styfen Sch\"ar, Bruno Sudret},
pdfsubject = {},
pdfkeywords = {NARX, Dynamical systems, Surrogate modeling, Autoregressive modeling, Principal component analysis, Least angle regression},
colorlinks = false,			
}

\usepackage[capitalize]{cleveref}							
\creflabelformat{equation}{#2(#1)#3}						
\crefrangelabelformat{equation}{#3(#1)#4 to #5(#2)#6}		
\labelcrefformat{subequation}{#2(#1)#3}						
\labelcrefrangeformat{subequation}{#3(#1)#4 to #5(#2)#6}	

\graphicspath{{figures}}

\begin{document}

\maketitle

\begin{abstract}
Time-variant reliability analysis is a critical task for ensuring the safety of engineering dynamical systems subjected to stochastic excitations. 
However, assessing failure probability for realistic systems with Monte-Carlo simulation-based methods is often computationally intractable due to the high cost of the underlying models and the large number of simulations required.
While surrogate models such as polynomial chaos expansions or Kriging are well-established for time-invariant reliability problems, their direct application to time-dependent systems remains challenging. This chapter introduces two advanced surrogate modeling frameworks designed specifically for dynamical systems: manifold-NARX (mNARX) and functional NARX ($\mathcal{F}$-NARX). \\
The mNARX approach constructs the surrogate on a reduced-order manifold of auxiliary state variables, enabling the efficient handling of high-dimensional inputs by embedding physical insight into a regression formulation. Conversely, the $\mathcal{F}$-NARX framework exploits the functional nature of system trajectories, extracting principal component features from continuous time windows to mitigate issues associated with discrete lag selection and long-memory effects. We demonstrate the efficacy of these methods on two benchmark reliability problems: a stochastic quarter-car model and a hysteretic Bouc-Wen oscillator. The results highlight that, when combined with suitably biased experimental designs, both frameworks accurately capture the tail behavior of the system response, enabling precise and efficient estimation of first-passage probabilities.
\end{abstract}

\section{Introduction}
\label{sec:Introduction}

Structural reliability analysis has reached a high level of maturity, providing rigorous frameworks to assess the safety of engineering systems under uncertainty. 
As detailed in previous chapters, the core task typically involves evaluating a limit state function $\vx \mapsto g(\vx, \cm(\vx))$, where $\vx$ denotes the vector of uncertain input parameters and $\cm$ represents the computational model of the system under consideration. 
When $\cm$ is computationally expensive, direct simulation methods such as Monte Carlo Simulation (MCS) become intractable, necessitating the use of \emph{surrogate models} (or metamodels) \citep{moustapha2022active}.

For time-invariant problems, where the quantity of interest (QoI) is a scalar or a low-dimensional vector (e.g., maximum displacement, total cost, or fatigue life), classical surrogate modeling techniques have become standard tools. Methods such as polynomial chaos expansions (PCE) \citep{Ghanembook2003,Xiu2002} and Kriging (or Gaussian Process modeling) \citep{Santner2003} allow for the accurate approximation of the map ${\vx \mapsto \cm(\ve{{x}})}$ with a limited computational budget. These techniques effectively replace the expensive solver $\cm$ within the reliability loop, enabling efficient estimation of failure probabilities \citep{Echard2011,MarelliSS2018,moustapha2022active}.

\subsection{The Challenge of Time-Dependent Reliability}

However, many critical engineering systems, such as offshore wind turbines, vehicle suspensions, or structures under seismic excitation, are inherently \emph{dynamical}. In these cases, the model output is not a scalar, but a time-dependent vector field or trajectory $y(t)$ defined over a time interval $[0, T_{\text{max}}]$. 
The reliability problem then transforms into a \emph{first-passage} problem, where one seeks the probability that the system trajectory exits a safe domain at least once during the observation window.

Applying classical surrogates to this class of problems presents fundamental challenges:

\begin{enumerate}
    \item \textbf{Evolving complexity over time:} In many dynamical systems, the complexity of the response increases with time due to nonlinear effects, accumulation of uncertainties, or interactions between system components.
    Standard surrogates struggle to capture this evolving complexity, especially for long-duration simulations \citep{le_2010_core,mai_2016,MaiSIAMUQ2017}.

    \item \textbf{High-dimensionality of the model response:} A direct ``pointwise'' application of classical surrogates would require training a separate model for every time step $t_i \in [0, T_{\text{max}}]$. For typical dynamical simulations with thousands of time steps, this approach is computationally prohibitive and ignores the strong temporal correlation between steps.
    
    \item \textbf{Inefficiency of projection methods:} A common workaround is to project the time-dependent output onto a functional basis, such as using principal component analysis (PCA) or discrete cosine expansions, and then surrogate the projection coefficients \citep{blatman2011principal,mai_2016}. While effective for smooth, linear dynamics, this approach often fails for highly nonlinear or path-dependent systems (e.g., those exhibiting hysteresis or abrupt transitions). To capture such complex features, a large number of basis functions is required, which in turn increases the dimensionality of the surrogate output, leading back to the problem of dimensionality.
    
    \item \textbf{Causality:} Projection-based methods treat the trajectory $y(t)$ as a monolithic functional object. They do not explicitly respect the principle of causality, \textit{i.e.} the fact that the state at time $t$ depends only on the history up to $t$. This can lead to nonphysical artifacts in the predicted trajectories, particularly in the presence of stochastic excitations.

\end{enumerate}

\subsection{Towards Time-Dependent Surrogates}

To address these limitations, it is necessary to move beyond functional approximation and adopt a \emph{dynamical} perspective. 
Instead of learning the map $\vx \mapsto y(t)$ directly, it is sometimes more effective to emulate the discrete-time evolution of the system. This leads to the class of Nonlinear AutoRegressive with eXogenous inputs (NARX) models \citep{billings_2013}.

NARX models construct the prediction at the current time step $t$ based on the immediate past history of the input and the output. By learning the \emph{recursive} law governing the system, rather than the full trajectory, NARX models naturally enforce causality and are independent of the total duration of the simulation.

\subsection{Chapter Outline}

This chapter introduces advanced strategies for time-variant reliability analysis using time-dependent surrogates. We focus on two recent extensions of the NARX framework designed to handle the complexities of modern engineering dynamics:
\begin{itemize}
    \item \textbf{Manifold NARX (mNARX):} A method that constructs the surrogate on a lower-dimensional manifold of auxiliary state variables, enabling the handling of complex, high-dimensional inputs (Section~\ref{sec:mNARX}).
    \item \textbf{Functional NARX ($\mathcal{F}$-NARX):} A functional approach that replaces discrete time lags with features extracted from continuous time windows, offering robustness against oversampling and long-memory effects (Section~\ref{sec:FNARX}).
\end{itemize}
We demonstrate how these methods can be integrated into a reliability framework to accurately estimate first-passage probabilities for highly nonlinear dynamical systems.
\section{Problem Statement: Reliability of Dynamical Systems}
\label{sec:ProblemStatement}

We consider a computational model of a dynamical system denoted by $\cm$. The system is subjected to a time-dependent random excitation $\vX(t)$ and governed by a set of parameters describing its physical properties. Let $\vXi$ be the random vector gathering all uncertainties in the problem, including both structural parameters and possibly excitation statistics, characterized by a joint probability density function (PDF) $\vXi \sim f_{\vXi}$. For a given realization of the random excitation $\vx(t)$, and of the uncertain input vector $\vxi$, the model output is a time-dependent response trajectory:
\begin{equation}
    y(t) = \cm(\vxi, \vx(\ct \leq t)),
\end{equation}
where the expression $\vx(\ct \leq t)$ denotes the history of the excitation up to time $t$. In a discretized setting, the full time axis $[0, T_\text{max}]$ is represented by $N_t$ time steps $\{t_0, t_1, \dots, t_{N_t-1}\}$ with a constant time step $\delta t$.

\subsection{Time-Variant Reliability}

The performance of the system is defined by a limit-state function $g(\vxi, \vx(t))$, which conventionally takes negative values when the system fails to meet a safety criterion. A common formulation involves a threshold $y_{\text{adm}}$ on the system response:
\begin{equation}\label{eqn:g(x)}
    g(\vxi, \vx(\ct \leq t)) = y_{\text{adm}} - y(\vxi, \vx(\ct \leq t)).
\end{equation}
While other formulations of the limit-state function are possible, the threshold-based definition is widely used in engineering applications, as it directly relates to admissible performance levels (e.g., maximum displacement, acceleration, or stress).

The \emph{instantaneous} probability of failure is the probability that the system is in a failure state at a specific time instant $t$. However, for dynamical systems, we are typically interested in the \emph{cumulative} probability of failure $P_{f,c}$, defined as the probability that a failure occurs \emph{at any time} during the interval $[0, T_\text{max}]$:
\begin{equation}\label{eqn:Pf_c_1}
    P_{f,c} = \mathbb{P}\left( \exists\, t \in [0, T_\text{max}] : g(\vXi, \vX(\ct \leq t)) \le 0 \right).
\end{equation}
In the discrete-time setting, this can be rewritten as the probability of the union of failure events over all time steps:
\begin{equation}\label{eqn:Pf_c_2}
    \begin{split}
    P_{f,c} \approx& \mathbb{P}\left( \bigcup_{k=0}^{N_t-1} \{ g(\vXi, \vX(\ct \leq t_k)) \le 0 \} \right)    \\
    =& \mathbb{P}\left( \min_{k \in \{0,\dots,N_t-1\}} g(\vXi, \vX(\ct \leq t_k)) \le 0 \right).
    \end{split}
\end{equation}
The event related to $k=0$ should be understood as the instantaneous initial failure $g(\vXi, \vX(t_0))\leq 0$.
Using the threshold condition in Eq.~\eqref{eqn:g(x)}, this is equivalent to finding the probability that the maximum response exceeds the admissible limit:
\begin{equation}\label{eq: first passage Pf}
    P_{f,c} = \mathbb{P}\left( y_{\text{adm}} - \max_{t \in \mathcal{T}} \cm(\vXi, \vX(\ct \leq t)) \le 0 \right).
\end{equation}

\subsection{Computational Challenges}

The standard estimator for $P_{f,c}$ is obtained via Monte Carlo Simulation (MCS). Let $\{\vxi^{(1)}, \dots, \vxi^{(N_{MCS})}\}$ be a set of independent realizations of the input vector, and $\{\vx^{(1)}(t), \dots, \vx^{(N_{MCS})}(t)\}$ be the corresponding realizations of the random input excitation. The estimator reads:
\begin{equation}
    \hat{P}_{f,c} = \frac{1}{N_{MCS}} \sum_{i=1}^{N_{MCS}} \mathbb{I}_{\mathcal{D}_f}(\vxi^{(i)},\vx^{(i)}(t)),
\end{equation}
where $\mathbb{I}_{\mathcal{D}_f}$ the failure indicator function, directly derived from the failure probability definition in Eq.~\ref{eq: first passage Pf}: $\mathbb{I}_{\mathcal{D}_f} = 1$ if $\max_{t \in \mathcal{T}} \cm(\vxi, \vx(\mathcal{T}\leq t)) > y_{\text{adm}}$, and $\mathbb{I}_{\mathcal{D}_f} = 0$ otherwise. Since structural failure is typically a rare event (with probabilities often in the range of $10^{-3}$ to $10^{-6}$), estimating $\hat{P}_{f,c}$ with sufficient accuracy requires a large number of simulations ($N_{MCS} \ge 10^5$). When each evaluation of $\cm(\vxi, \vx(t))$ involves solving complex differential equations (e.g., nonlinear, transient finite element analysis), the total computational cost becomes prohibitive. 
Based on Eq.~\eqref{eq: first passage Pf}, a surrogate model of the maximum response $\vxi \mapsto \widehat\cm_{\max}(\vxi)\approx \max\limits_{\vX(t)} \cm(\vxi,\vx(t))$ could be envisaged.
However, such a construction is difficult and can be rather inaccurate, since the max function introduces nonsmoothness in the mapping of parameters $\vxi$ to the response, even though it has seen successful application in different fields (see, {\it e.g.}, \cite{Wu2022}).

Several tools in the reliability analysis literature have been developed to improve the efficiency of simulation-based methods for time-dependent problems, with varaince reduction methods such as subset simulation \citep{Du2019}, importance sampling \citep{wang2021importance}, or the probability density evolution method \citep{li2004probability}.
Nonetheless, the literature on surrogate-based reliability analysis of time-dependent systems is far less extensive, mostly limited to the use of Gaussian processes \citep{hu2016single}, response surfaces \citep{Zhang2017_Reliability}, and first-order approximations \citep{hu2015first}.

\section{Surrogate modeling for Time-Dependent Problems and the NARX Framework}
\label{sec:NARX_basics}

A different approach to building a surrogate model for reliability analysis of time-dependent problems is to directly approximate the entire time-dependent response of the computational model:
\begin{equation}
\label{eq:time dependent surrogate}
\widehat y(t) = \widehat\cm(\vxi, \vx(\ct \leq t)) \approx \cm(\vxi, \vx(\ct \leq t)),
\end{equation} 
where $\vx(\ct \leq t)$ explicitly gathers the $M$ time-dependent uncertain inputs to the system up to time $t$.
To construct such surrogates for dynamical systems, we rely on the Nonlinear AutoRegressive with eXogenous inputs (NARX) framework. Unlike time-invariant surrogates that map inputs directly to outputs, NARX models aim at approximating the entire dynamical evolution of the system \citep{billings_2013}.

\subsection{Standard NARX Formulation}

A standard NARX model assumes that the current output $y(t)$ depends on the past values of the output (autoregressive terms) and the current and past values of the input (exogenous terms). Formally, the approximation $\hat{y}(t)$ is given by:
\begin{equation}\label{eq:NARX}
    \hat{y}(t) = \mathcal{F}\Big( y(t-\delta t), \dots, y(t-n_y \delta t), \vx(t), \dots, \vx(t-n_x \delta t), \vxi \Big) + \epsilon,
\end{equation}
where $n_y$ and $n_x$ represent the maximum time lags for the output and input, respectively, and $\epsilon$ is the residual error. The vector gathering the delayed inputs and outputs is referred to as the \emph{regressor vector}:
\begin{equation}\label{eq:Phi}
    \ve{\phi}(t) = \left[ y(t-\delta t), \dots, y(t-n_y \delta t), \vx(t), \dots, \vx(t-n_x \delta t),\vxi \right]^\top.
\end{equation}

\subsection{Polynomial NARX and Sparse Regression}

The mapping function $\mathcal{F}$ can take various forms, such as neural networks, Gaussian processes, or polynomial expansions. In this work, we focus on \emph{polynomial NARX} models, where $\mathcal{F}$ is expanded onto a basis of multivariate polynomials. The output at time $t$ is approximated as:
\begin{equation}
    \hat{y}(t) = \sum_{j=0}^{P-1} \theta_j \Psi_j(\ve{\phi}(t)),
    \label{eq:PNARX}
\end{equation}
where $\{\Psi_j\}$ are polynomial functions ({\it e.g.}, products of monomials) constructed from the regressors in Eq.~\eqref{eq:Phi}, and $\ve{\theta} = \{\theta_0, \dots, \theta_{P-1}\}$ represent a set of unknown coefficients.

Calibrating the autoregressive model involves estimating $\ve{\theta}$ from a set of training trajectories. In polynomial NARX, this is usually cast as a linear regression problem with regression matrix $\ve{\Psi} \in \mathbb{R}^{\tilde{N} \times P}$:
\begin{equation}\centering
\label{eq:Phi_matrix}
        \ve{\Psi} = \begin{pmatrix}
            \Psi_1\left(\ve{\phi}(t_{\min})\right), & ~~\dots,~~ & \Psi_P\left(\ve{\phi}(t_{\min})\right) \\
            \Psi_1\left(\ve{\phi}(t_{\min}+\delta t)\right), & ~~\dots,~~ & \Psi_P\left(\ve{\phi}(t_{\min}+\delta t)\right) \\
            \vdots & \ddots & \vdots \\
            \Psi_1\left(\ve{\phi}(t_{\min}+(N-1)\delta t)\right), & ~~\dots,~~ & \Psi_P\left(\ve{\phi}(t_{\min}+(N-1)\delta t)\right)
    \end{pmatrix}
\end{equation}
 and output vector $\ve{y} \in \mathbb{R}^{\tilde{N}}$:
\begin{equation}
        \ve{y} = \begin{pmatrix}
            y(t_{\min}) \\
            y(t_{\min}+\delta t) \\
            \vdots \\
            y(t_{\min}+(N-1)\delta t)
        \end{pmatrix}
\end{equation}
where $t_{\min} = \max(n_y, n_x) \delta t$, and $\tilde{N} = N - \max(n_y, n_x)$.\\
Regression matrices and output vectors from {\it multiple traces} can also be concatenated:
\begin{equation}\label{eq:large_Phi_matrix}
    \ve{\Psi}_\text{ED} = \begin{pmatrix}
        \ve{\Psi}^{(1)} \\
        \vdots \\ 
        \ve{\Psi}^{(N_\text{ED})}
    \end{pmatrix}
    \quad
    \ve{y}_\text{ED} = \begin{pmatrix}
            \ve{y}^{(1)} \\ 
            \vdots \\
            \ve{y}^{(N_\text{ED})} 
    \end{pmatrix}.
\end{equation}

In case of long traces (thousands of timesteps), or large experimental designs, the number of total time steps can become too large. 
In this case {\it subsampling} can be used:
\begin{equation}
    \ve{\Psi}_S = \begin{pmatrix}
        \ve{\Psi}_{\text{ED}, r_1} \\
        \ve{\Psi}_{\text{ED}, r_2} \\
        \vdots \\
        \ve{\Psi}_{\text{ED}, r_k}
    \end{pmatrix}
    \quad
    \ve{y}_S = \begin{pmatrix}
        \ve{y}_{\text{ED}, r_1} \\
        \ve{y}_{\text{ED}, r_2} \\
        \vdots \\
        \ve{y}_{\text{ED}, r_k}
    \end{pmatrix}
\end{equation}
where $r_i \in \{1, 2, \dots, |\ve{y}_\text{ED}|\}$ are randomly or deterministically drawn.
The coefficient vector $\ve{\theta}$ is then estimated by solving the least-squares problem:
\begin{equation}\label{eqn:OLS}
    \hat{\ve{\theta}} = \arg\min_{\ve{\theta}} \| \ve{y}_S - \ve{\Psi}_S \ve{\theta} \|_2^2.
\end{equation}
In many cases, the number of candidate basis functions $P$ can become intractably large, especially when considering high polynomial degrees or significant interaction terms ({\it i.e.} multivariate $\Psi_j$'s in Eq.~\eqref{eq:PNARX}). 
To avoid overfitting and maintain computational efficiency, we employ \emph{sparse regression} techniques, such as regularized regression (LASSO) \citep{Tibshirani_1996_LASSO}, which promotes sparsity in the coefficient vector by adding an $\ell_1$-norm penalty to the least-squares objective:
\begin{equation}\label{eqn:LASSO}
    \hat{\ve{\theta}} = \arg\min_{\ve{\theta}} \| \ve{y}_S - \ve{\Psi}_S \ve{\theta} \|_2^2 + \gamma \|\ve{\theta}\|_1.
\end{equation}
To solve the latter, we use least angle regression (LARS) \citep{Efron_2004}, which efficiently selects the most significant basis functions, while simultaneously estimating both their coefficients and the regularization parameter $\gamma$. This results in a sparse coefficient vector $\hat{\ve{\theta}}$ where most entries are zero, yielding a parsimonious model that generalizes well to unseen data \citep{mai_2016}.

\subsection{Limitations of Standard NARX}

While standard polynomial NARX models are effective for many systems, they exhibit limitations when applied to the complex reliability problems described in Section~\ref{sec:ProblemStatement}:
\begin{itemize}
    \item \textbf{Curse of dimensionality:} If the input $\vx(t)$ is high-dimensional (e.g., a spatial field), the size of the regressor vector $\ve{\phi}(t)$ exponentially increases, making standard regression infeasible.
    \item \textbf{Lag Selection:} Identifying the optimal lags $n_y$ and $n_x$ is nontrivial. Inappropriate choices can lead to models that fail to capture the system's memory, often resulting in instability over long periods of time.
\end{itemize}
These challenges motivate the development of the advanced architectures presented in the following sections: manifold NARX (mNARX) \citep{schaer_2024MSSP} and functional feature-based NARX ($\mathcal{F}$-NARX) \citep{schar2025FNARX}.
\section{Method I: Manifold NARX (mNARX)}
\label{sec:mNARX}

Standard NARX models \citep{billings_2013} provide a robust framework for black-box modeling of dynamical systems. However, they face significant challenges when applied to complex engineering problems where the mapping from the exogenous input $\vx(t)$ to the quantity of interest (QoI) $y(t)$ is highly nonlinear, nonsmooth, or involves high-dimensional excitations (e.g., spatial wind or earthquake fields). To address these limitations, the \emph{manifold-NARX} (mNARX) framework was introduced by \cite{schaer_2024MSSP}.

The core rationale of mNARX is that the complexity of the input/output mapping ${\vx(t) \mapsto y(t)}$ can be significantly reduced if the surrogate model is constructed on a more informative {\it manifold}, $\ve{\zeta}(t)$, rather than directly on the original input space.
This manifold is constructed by concatenating a reduced representation of the exogenous input, and a set of auxiliary state variables that capture the internal dynamics of the system, resulting in a more informative input space for the surrogate model.

\subsection{Methodological Framework}

The mNARX approach constructs a multistep surrogate model through three distinct phases: input preprocessing, manifold construction, and surrogate training.

\subsubsection{Input Preprocessing\\} 
Let the input be denoted by $\vx(t) \in \mathbb{R}^{M}$, where $M = O(10^2-10^4)$. Direct use of $\vx(t)$ in a NARX regression is computationally intractable due to the curse of dimensionality.

\begin{figure}[tb]
    \centering
    \includegraphics[width=0.85\linewidth,clip,trim=0 35 0 0]{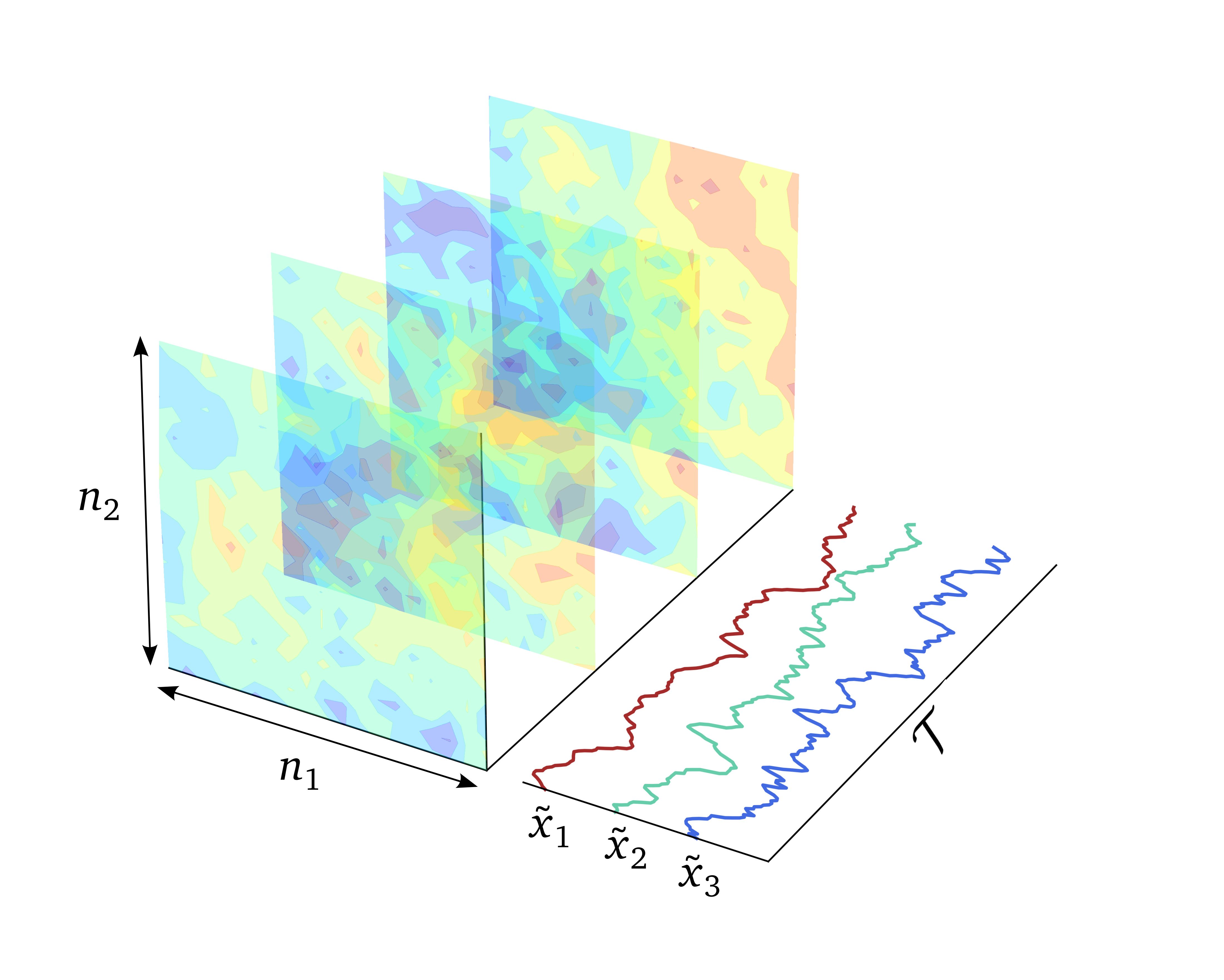}
    \caption{Graphical illustration of the input preprocessing step in mNARX, where a high-dimensional input $\vx(t)$ (\emph{e.g.} a grid over the $(y,z)$ coordinates) is projected onto a lower-dimensional representation $\tilde{\vx}(t)$ using dimensionality reduction techniques (image adapted from \cite{schaer_2024MSSP}).
    }
    \label{fig:mNARX preproc}
\end{figure}

The first step is therefore to project the input onto a lower-dimensional space using dimensionality reduction techniques suitable for the physics of the problem, such as principal component analysis (PCA) \citep{Jolliffe2002} or discrete cosine transforms \citep{rao1990discrete}. This yields a reduced input vector $\tilde{\vx}(t) \in \mathbb{R}^{m}$ with $m \ll M$:
\begin{equation}
    \tilde{\vx}(t) = \mathcal{G}(\vx(t)),
\end{equation}
where $\mathcal{G}$ denotes the compression mapping. This step preserves the temporal dimension while reducing the spatial complexity.
This process is represented graphically in Figure~\ref{fig:mNARX preproc}.

\subsubsection{Manifold Construction\\}

\begin{figure}[tb]
    \centering
    \includegraphics[width=0.7\linewidth]{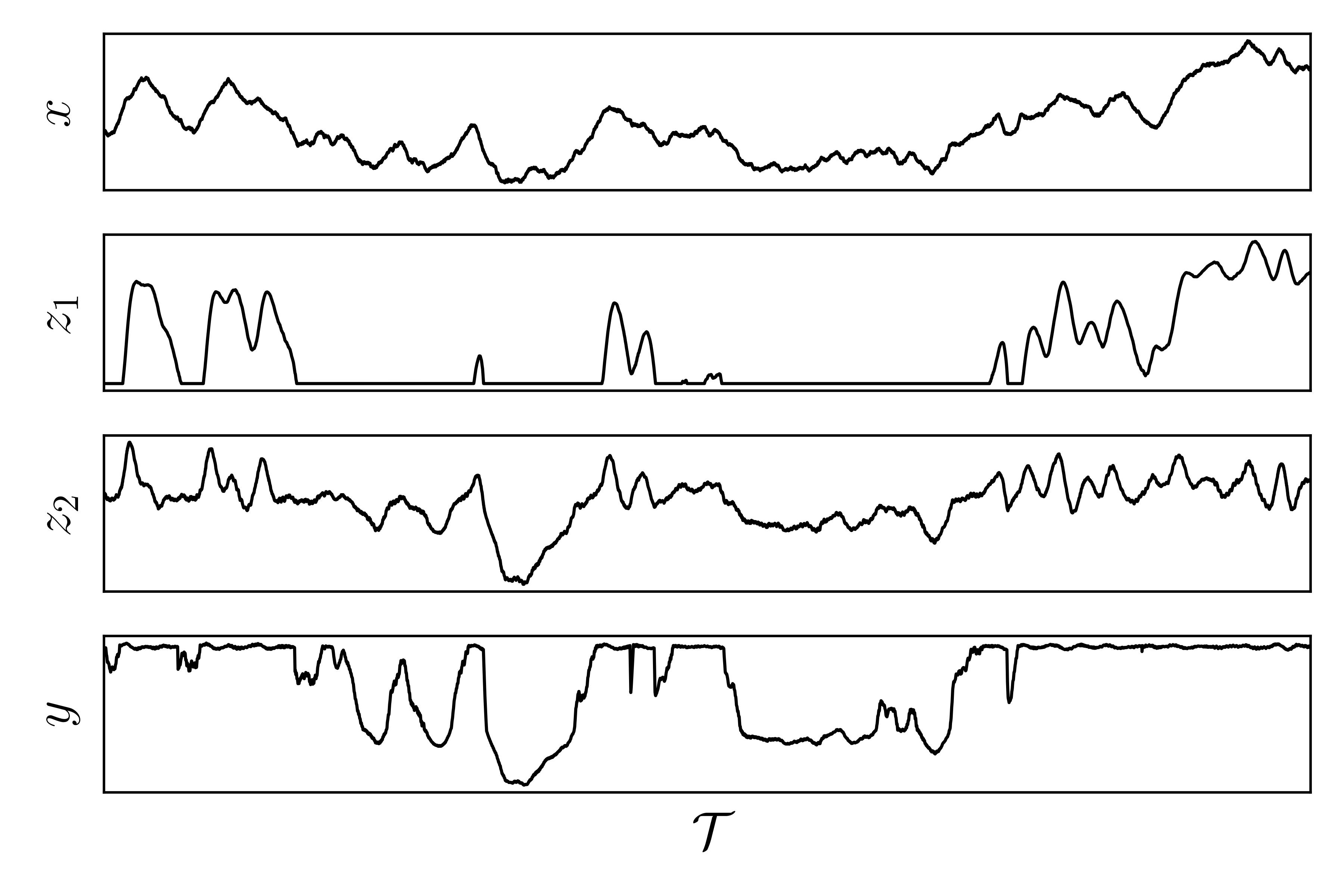}
    \caption{Graphical illustration of the manifold construction in mNARX, where auxiliary quantities $\ve{z}(t)$ are sequentially predicted to form the manifold $\ve{\zeta}(t)$ used for predicting the QoI $y(t)$ (image adapted from \cite{schaer_2024MSSP}).
    }
    \label{fig:mNARX aux}
\end{figure}

The central innovation of mNARX is the embedding of physical insight into the surrogate via \emph{auxiliary quantities} \citep{schaer_2024MSSP}. The manifold $\ve{\zeta}(t)$ is defined as the concatenation of the reduced input and a set of $n_z$ auxiliary state variables $\ve{z}(t) = \{z_1(t), \dots, z_{n_z}(t)\}^\top$:
\begin{equation}
    \ve{\zeta}(t) = \left\{ \tilde{\vx}(t), \ve{z}(t) \right\}.
\end{equation}
These auxiliary quantities are intermediate system responses that ``unfold'' the nonlinear dynamics. They are constructed incrementally: the first quantity $z_1$ is predicted based on the input $\tilde\vx(t)$ only, the second $z_2$ based on the input $\tilde\vx$ and $z_1$, and so forth. Formally, for $k = 1, \dots, n_z$:
\begin{equation}
    z_k(t) \approx \widehat{\cm}_k \left( \tilde{\vx}(\ct \leq t), z_1(\ct < t), \dots, z_{k-1}(\ct < t) \right).
\end{equation}
A graphical illustration of this process is provided in Figure~\ref{fig:mNARX aux}.
This hierarchical structure allows the surrogate to discover the internal causal chain of the physical solver. Finally, the quantity of interest $y(t)$ is predicted using the full manifold:
\begin{equation}
    y(t) \approx \widehat{\cm}_{y} \left( \ve{\zeta}(\ct \leq t) \right).
\end{equation}

\subsubsection{Training the mNARX model\label{sec:mNARX training}\\}

The individual mappings $\widehat{\cm}_k$ and $\widehat{\cm}_y$ are trained using the polynomial NARX formulation in Section~\ref{sec:NARX_basics}. To handle the potentially large number of candidate terms in the polynomial expansion, we use least angle regression (LARS) \citep{Efron_2004} to obtain a sparse set of NARX coefficients.

A critical challenge in time-variant reliability is that failure events are rare and driven by extreme excitations. As a result, standard random sampling of training trajectories may fail to cover the ``tail'' regions of the response distribution.

To overcome this, we employ a \emph{biased experimental design} approach. 
Instead of randomly sampling realizations purely from the input distribution $f_{\vX}$, we select input excitations that are more likely to induce extreme system responses. 
Because the exact strategy to obtain such samples is problem-dependent, we detail it directly in Section~\ref{sec:mNARX biased ED}.

\subsection{Application: Quarter-Car Model Subject to Stochastic Excitation}
 
\begin{figure}[tb]
    \centering
    \begin{minipage}{.4\linewidth}
        \centering
        \includegraphics[width=.6\linewidth]{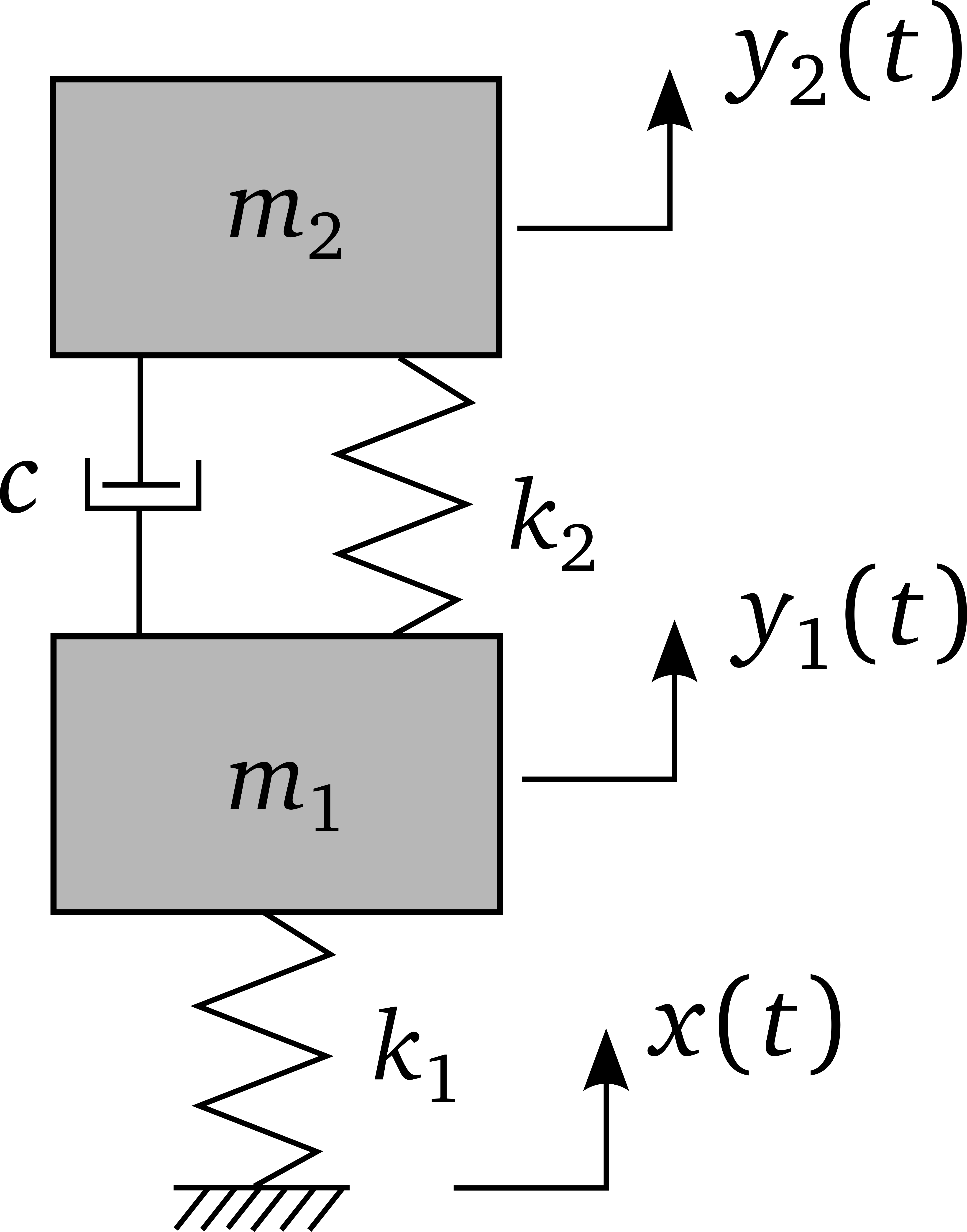}    
    \end{minipage}
    \begin{minipage}{.55\linewidth}
        \centering
        \begin{tabular}{@{}lcc@{}}
            \hline
            Parameter & Unit & Value \\ \hline
            Spring stiffness $k_1$ & N/mm & $5{,}000$ \\
            Spring stiffness $k_2$ & N/mm & $1{,}000$ \\
            Mass $m_1$ & kg & $50$ \\ 
            Mass $m_2$ & kg & $10$ \\
            Damping ratio  $c$ & Ns/mm & $50$ \\ \hline
        \end{tabular}
    \end{minipage}
    
    \caption{Schematic of the Quarter-Car model subjected to stochastic road excitation $x(t)$, and corresponding parameters
    .}
    \label{fig:QuarterCar}
\end{figure}

To illustrate the mNARX capability in handling reliability problems, we consider the stochastic quarter-car model shown in Figure~\ref{fig:QuarterCar}. 
This classical benchmark represents a two-degree-of-freedom system governed by coupled nonlinear differential equations:
\begin{equation}
    \begin{cases}
        m_2 \ddot{y}_2(t) = -k_2(y_2(t) - y_1(t))^3 - c(\dot{y}_2(t) - \dot{y}_1(t)) \\
        m_1 \ddot{y}_1(t) = k_2(y_2(t) - y_1(t))^3 + c(\dot{y}_2(t) - \dot{y}_1(t)) + k_1(x(t) - y_1(t))
    \end{cases}
    \label{eq:quarter_car}
\end{equation}
where $y_1(t)$ and $y_2(t)$ are the displacements of the wheel and vehicle body (chassis), respectively, and $x(t)$ is the stochastic road excitation. The nonlinearity arises from the cubic stiffness $k_2$ of the suspension spring. 
The model parameters are summarized in the Table in Figure~\ref{fig:QuarterCar}.

For generality, the system is subject to a stochastic excitation given by the following expression:
\begin{equation}\label{eq:mass_spring_system_exo_input}
        x(t) = \frac{1}{N_\omega}\sum_{i=1}^{N_\omega} A_i \sin \left( 2\pi B_i t + C_i\right),
\end{equation}
where $N_\omega$ is a discrete uniform random variable with $\mathbb P(N_\omega = i) = 1/5$ for $i=1, \dots, 5$, $A_i$, $B_i$ and $C_i$ are independent uniform random variables in the interval $[-1,1]$. The excitation is simulated over the time interval $t \in [0, T_\text{max}]$ with $T = 30$ s.

The goal is to compute the probability that the upper mass $m_2$ exceeds a prescribed threshold $y_{\text{adm}}$, {\it i.e.}, $P_f = \mathbb{P}(\max_{t \in [0,T_\text{max}]} |y_2(t)| > y_{\text{adm}})$.
This example is particularly challenging for classical approaches both because of its nonlinearity, and because of the complexity of the input excitation in Eq.~\eqref{eq:mass_spring_system_exo_input}, which is a superposition of multiple harmonics with random amplitudes, frequencies, and phases.

\subsection{Generating a biased experimental design}
\label{sec:mNARX biased ED}

\begin{figure}\centering
    \includegraphics[width=0.48\linewidth]{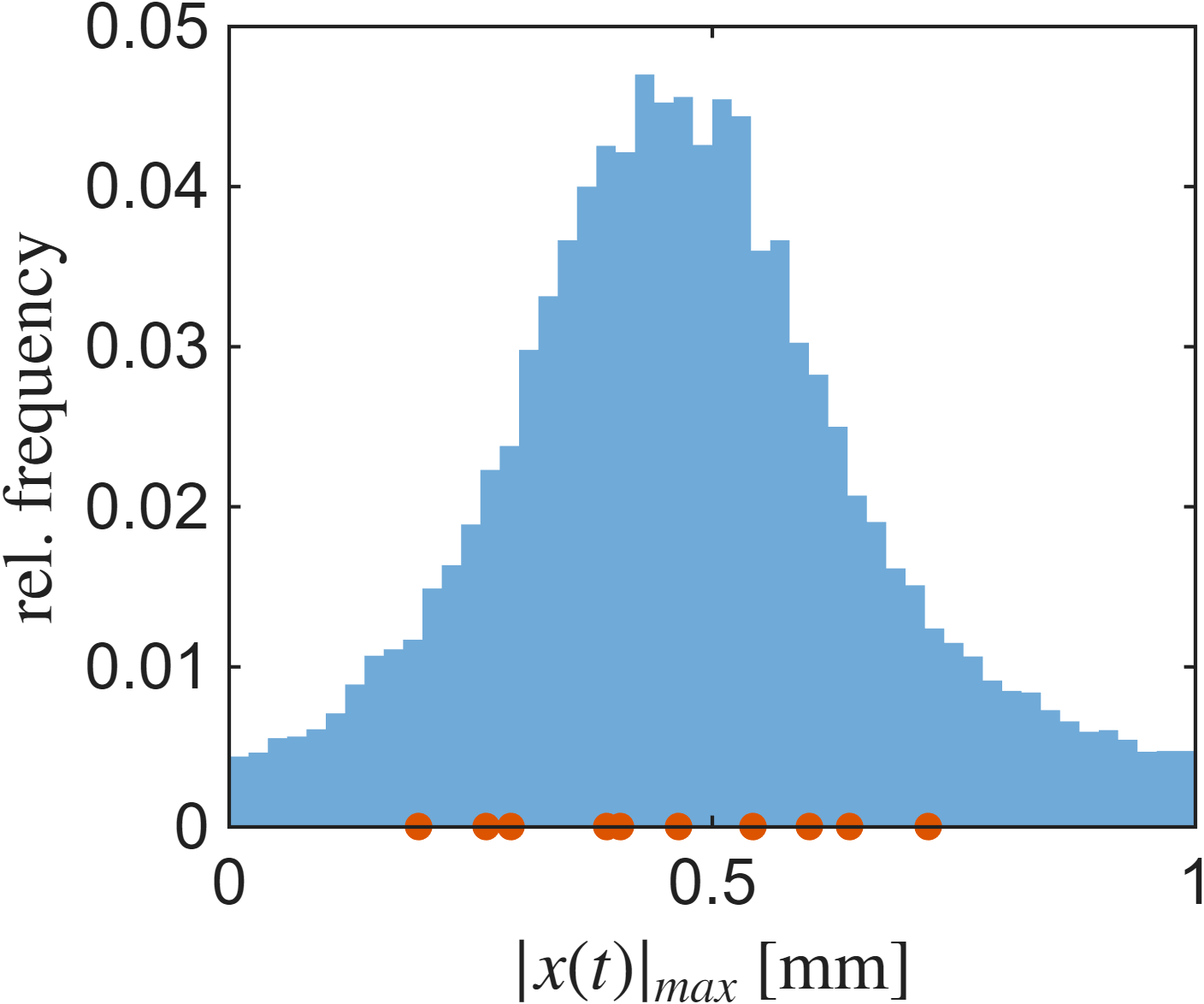}
    \includegraphics[width=0.48\linewidth]{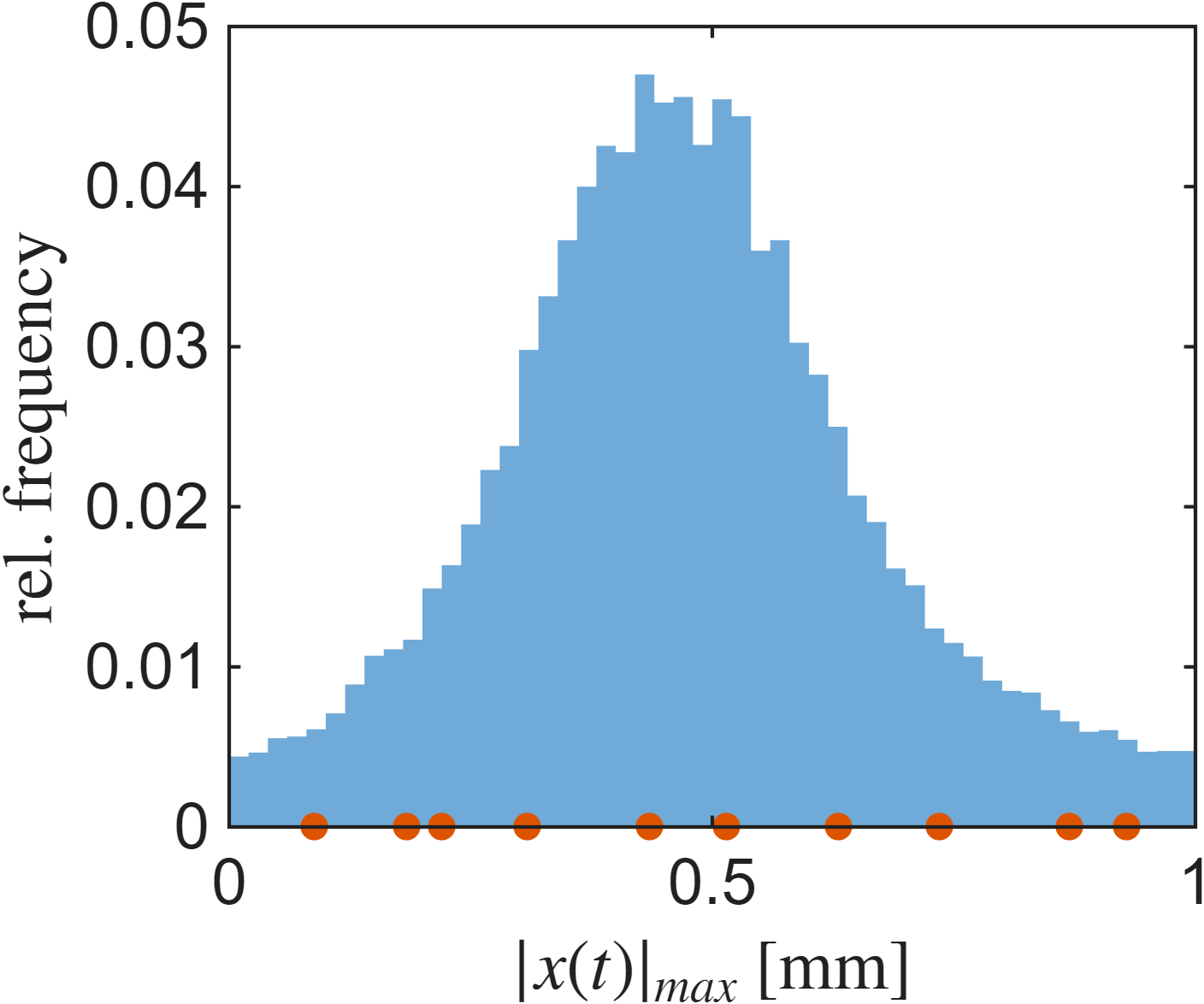}
    \caption{Graphical depiction of the biased sampling strategy in Section~\ref{sec:mNARX biased ED}. In both panels, the histogram shows the distribution of maximum absolute excitation amplitudes $|x(t)^{(i)}|_{\max}$ for a large candidate set of $2\cdot10^4$ input realizations, while the red dots represent the selected amplitudes. Left: standard random subsampling; Right: the biased selection ensures a much wider coverage of extreme amplitudes.}
    \label{fig:mNARX biased ED}
\end{figure}

To generate an experimental design that covers the largest range of model behaviors, we bias the experimental design by generating a large set of candidate input excitations, selecting a subset of them based on their maximum absolute amplitude $|x(t)^{(i)}|_{\max}$. 
This approach is showcased in Figure~\ref{fig:mNARX biased ED}, where we gather the maximum absolute excitation amplitudes of the candidate set in a histogram, and compare simple subsampling with the biased selection strategy.
More in detail, we proceed as follows: we generate a large sample of $N_c = 2\cdot10^4$ candidate input realizations $\{x(t)^{(i)}\}_{i=1}^{N_c}$ from Eq.~\eqref{eq:mass_spring_system_exo_input}. We then identify the range of maximum absolute amplitudes \begin{equation}
    A_{\min} = \min_{i=1,\dots,N_c} |x(t)^{(i)}|_{\max}, \quad A_{\max} = \max_{i=1,\dots,N_c} |x(t)^{(i)}|_{\max}.
\end{equation}
Finally, we create a random sample of $N_\text{ED}$ amplitudes $\{A_j\}_{j=1}^{N_\text{ED}}$ uniformly distributed in $[A_{\min}, A_{\max}]$. For each amplitude $A_j$, we then select the candidate realization $x(t)^{(i^*)}$ that is closest to $A_j$.
Note that at this stage, no nonlinear transient simulation of the system has been run yet.

Once the experimental design is generated (the size $N_\text{ED}$ of which will be specified directly in the applications below), we evaluate the computational model for each selected input realization to obtain the training data for the mNARX surrogate.

\subsection{mNARX model selection configuration and validation process}

The mNARX model is configured with $y_1(t)$ (wheel displacement) as the auxiliary quantity $z_1(t)$ for predicting the QoI $y_2(t)$. 

Once the experimental design is selected and the training data are generated, 
an mNARX model configuration must be defined.
To achieve this, we perform a preliminary analysis based on the approach described in \cite{mai_2016}:
\begin{enumerate}
    \item We select one single training trace from the experimental design, namely the one with the maximal response amplitude $\max_{t \in [0,T_\text{max}]} |y_2(t)|$.
    \item We then train several sparse candidate mNARX models with LARS on this single trace, varying hyperparameters such as the maximum polynomial degree, interaction terms, lags, etc.
    \item For each set of hyperparameters, we re-calculate the coefficients of the selected basis terms using ordinary least squares (Eq.~\eqref{eqn:OLS}) and the full experimental design, and evaluate the mean forecast error as:
    \begin{equation}\label{eq:lars_mean_forecast_error}
            \overline{\varepsilon} = \frac{1}{N_\text{ED}} \sum_{i=1}^{N_\text{ED}} \varepsilon^{(i)}
            \quad\text{with}\quad
            \varepsilon^{(i)} = \frac{1}{N} \frac{
                \displaystyle 
                \sum_{j=0}^{N-1} \left( y^{(i)}(j\delta t) - \hat{y}^{(i)}(j\delta t) \right)^2
            }{
                \text{Var}(\vy^{(i)}) + \gamma
            },
    \end{equation}
    where $\gamma$ is a small regularization constant to avoid division by zero.
    \item The mNARX model with the lowest mean forecast error $\overline{\varepsilon}$ is selected as the final configuration.
    \item  This process is repeated for each auxiliary quantity and QoI.
\end{enumerate} 

The final configuration of mNARX is reported in Table~\ref{tab:mNARX quarter car config}.
Additionally, we show the configuration of a classical NARX model that directly predicts the QoI $y_2(t)$ solely from $x(t)$.
\begin{table}[tb]
    \centering
    \caption{Final mNARX configuration for the Quarter-Car model. The lags are given in number of time steps, with $\delta t = 0.01$ s.}
    \label{tab:mNARX quarter car config}
    \begin{tabular}{lccc}
        \hline
        Property & mNARX for $\widehat y_1~~~~~$ & mNARX for $\widehat y_2~~~~~$ & NARX for $\widehat y_2$ \\ \hline
        Exogenous inputs & $x(t)$ & $x(t), \widehat y_1(t)$ & $x(t)$ \\
        Exogenous input lags & {$0$} & {$0$}, {$0$} & {$0$} \\
        Autoregressive lags & {$\delta t, 2\delta t$}, & {$\delta t, 2\delta t$} & {$\delta t, 2\delta t$} \\
        Maximum Polynomial degree & 1 & 3 & 3 \\
        Interaction order & 1 & 2 & 2 \\\hline
    \end{tabular}

\end{table}

\subsection{mNARX performance and convergence with $N_\text{ED}$}
To assess the performance of this methodology as a function of the available model evaluations, we perform the analysis on three different experimental design sizes: $N_\text{ED} = \{10, 50, 100\}$. 
For each size, we compare the performance of classical subsampling and biased sampling on an out-of-sample validation set of size $N_{\text{val}} = 2\cdot10^4$.
\begin{figure}[tb]
        \begin{minipage}{0.49\linewidth}
        \centering
        \includegraphics[width=\linewidth]{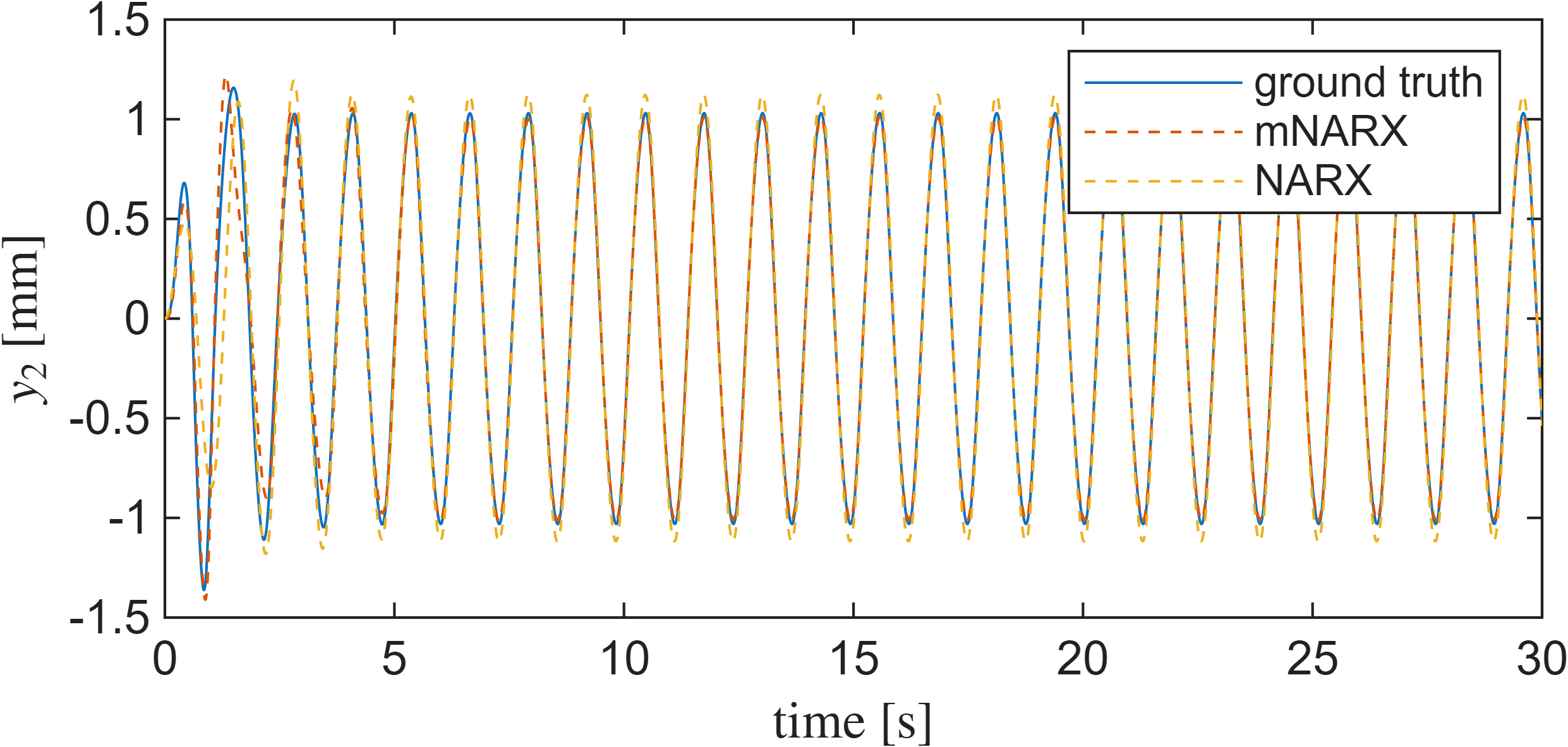}
        \includegraphics[width=\linewidth]{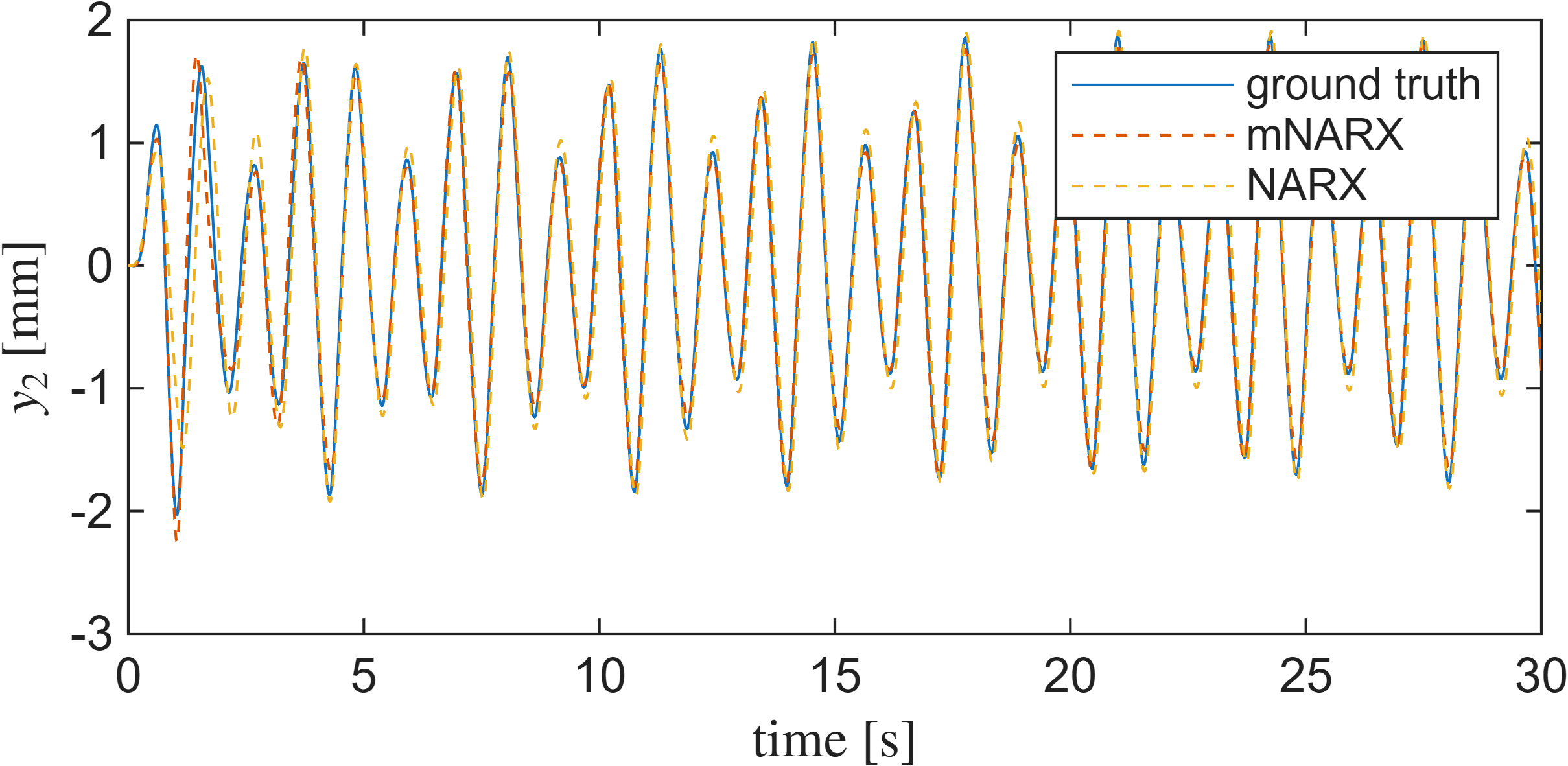}
        \includegraphics[width=\linewidth]{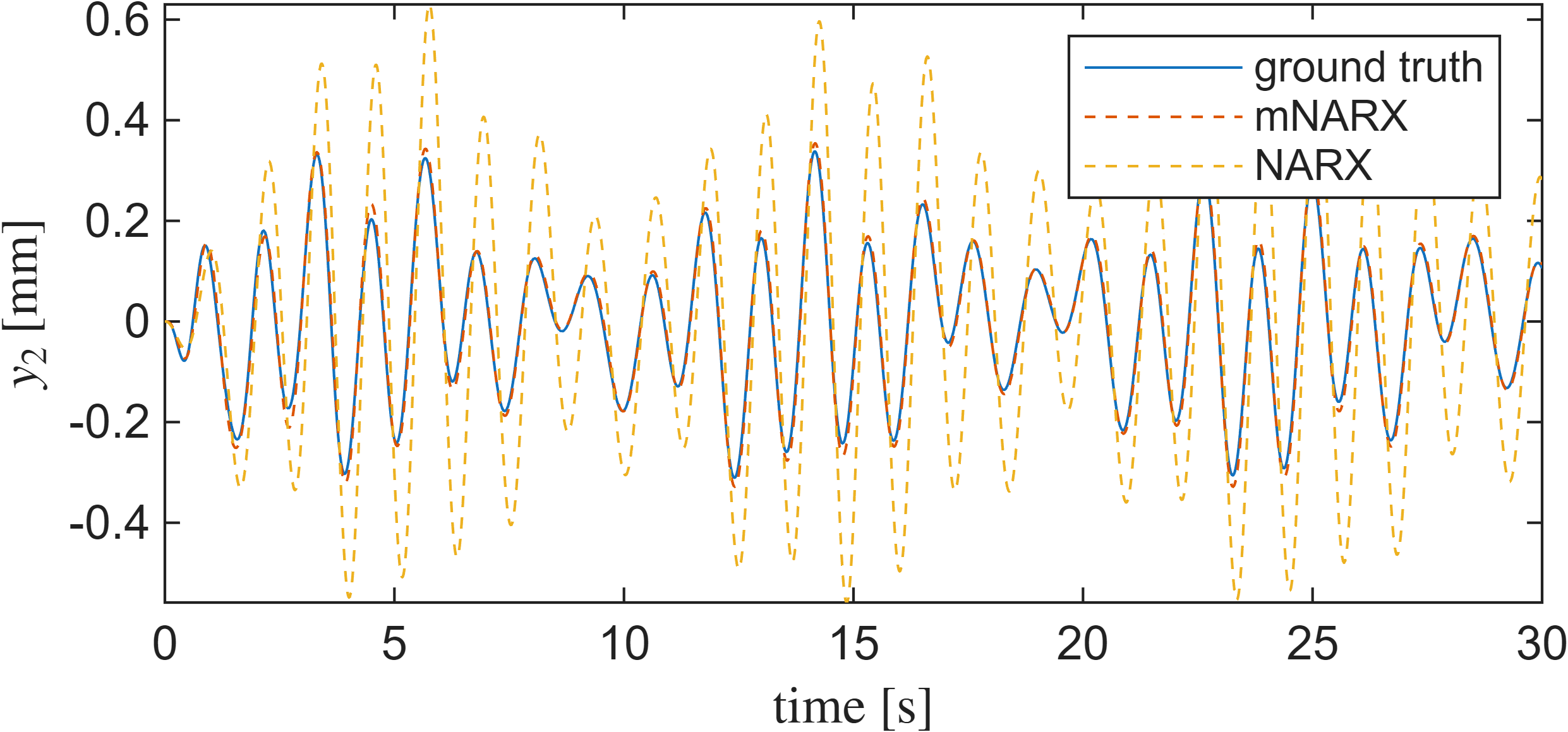}
    \end{minipage}
    \begin{minipage}{0.49\linewidth}
        \centering
        \includegraphics[width=\linewidth]{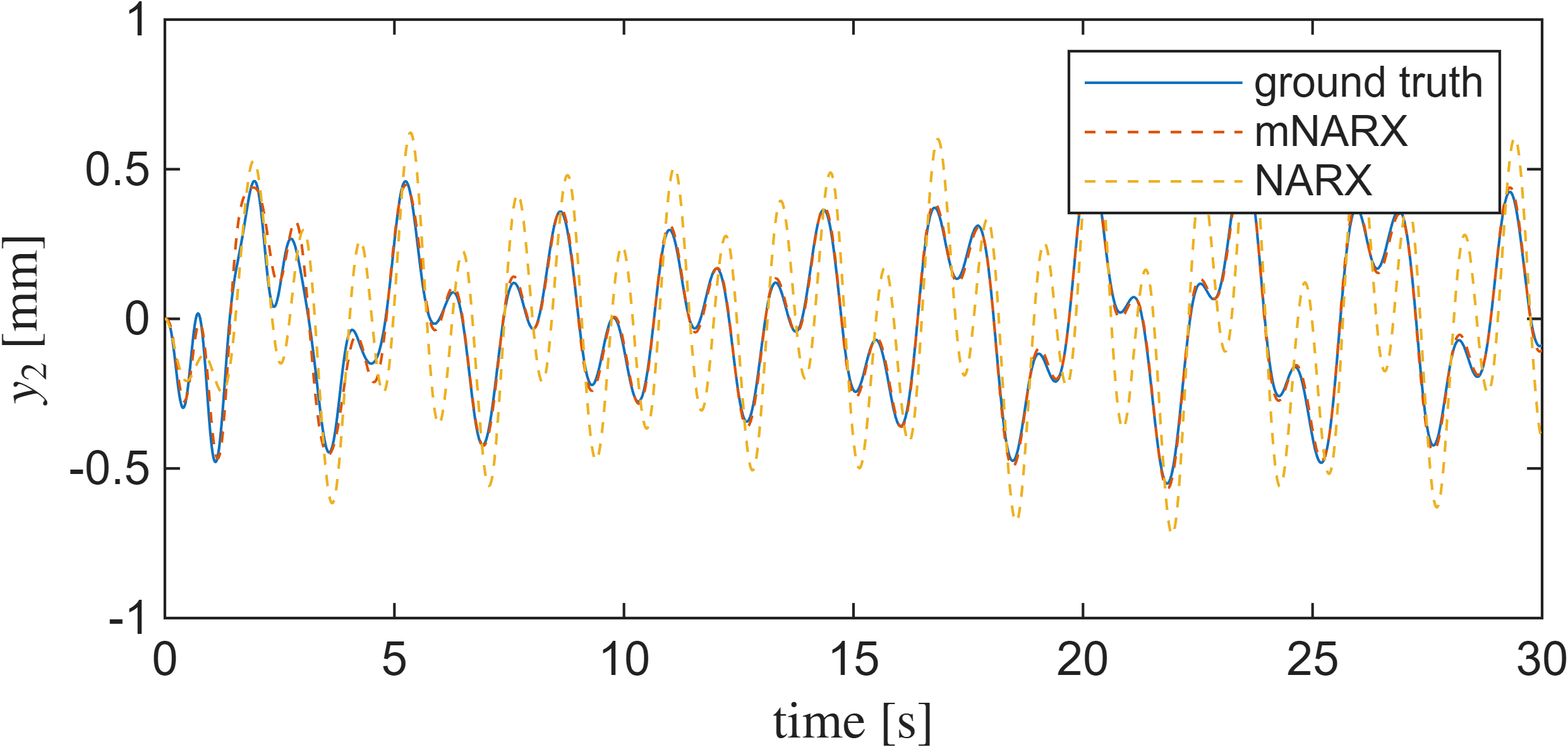}
        \includegraphics[width=\linewidth]{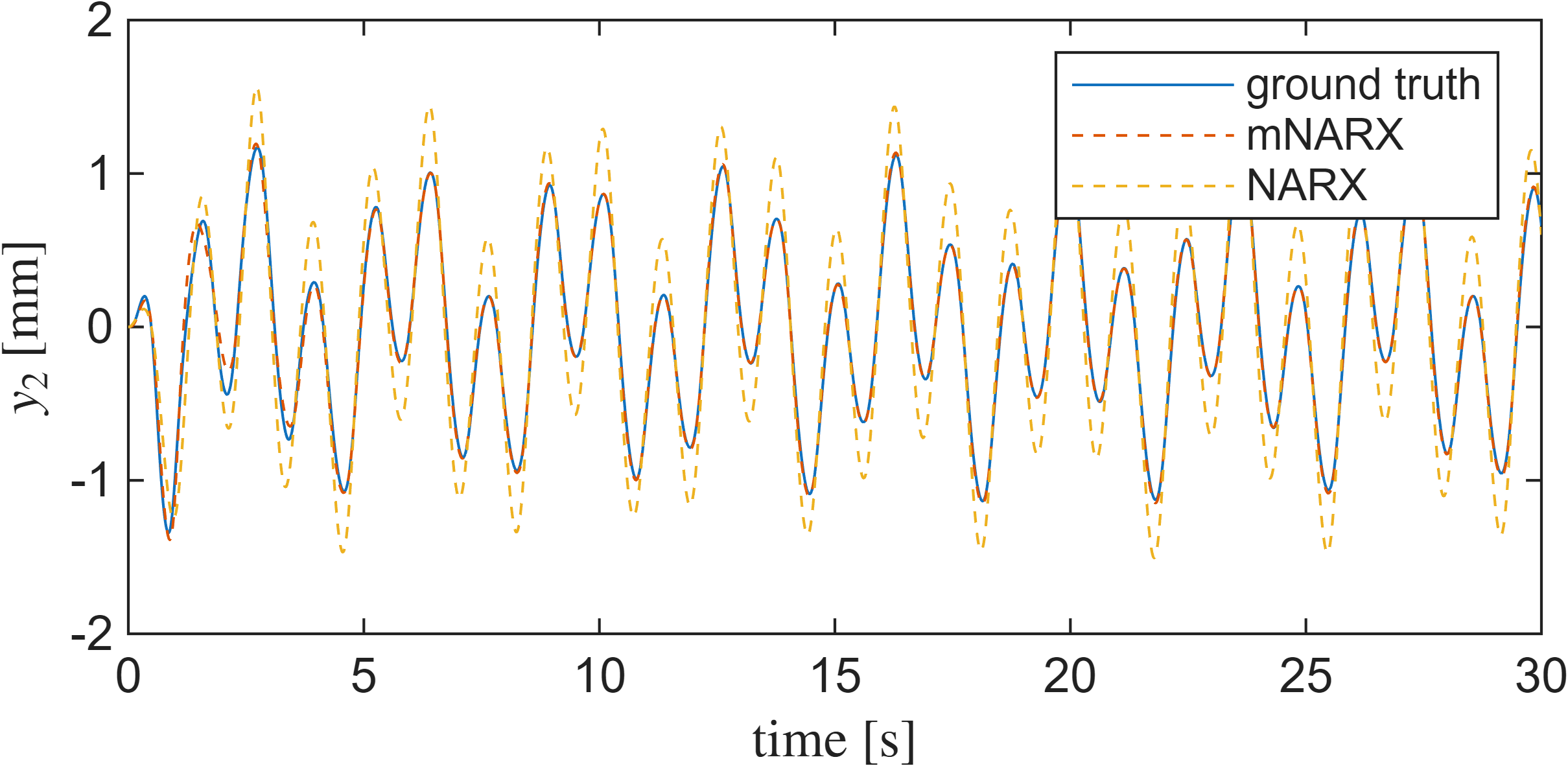}
        \includegraphics[width=\linewidth]{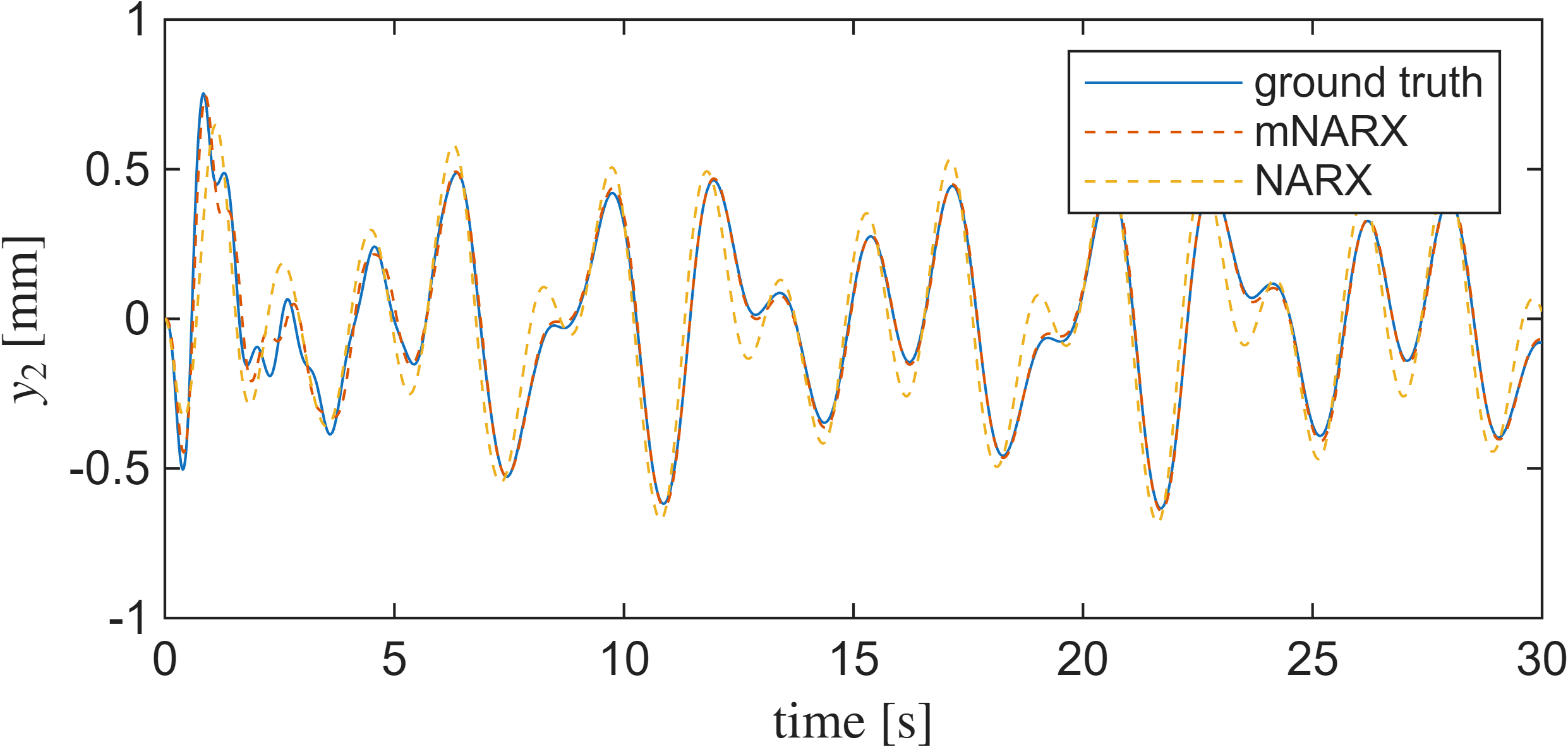}
    \end{minipage}
    \caption{Selection of out-of-sample response traces for an mNARX model and a classical NARX model trained on $N_\text{ED} = 50$ samples generated with the unbiased sampling strategy.}
    \label{fig:mNARX quarter car traces}
\end{figure}

We begin by comparing in Figure~\ref{fig:mNARX quarter car traces} a random selection of real and approximated full time series obtained with unbiased sampling on $N_\text{ED} = 50$. 
These results demonstrate that the mNARX surrogate can accurately capture the system dynamics across a wide range of behaviors, from mild to extreme responses, even with extremely limited data, and without the use of the biased sampling strategy.
In comparison, the classical NARX model performs significantly worse and shows a clear deviation from the ground truth.

\begin{figure}[tb]
    \centering
       \begin{minipage}{.08\linewidth}
        \centering\hspace{-1cm}
        \rotatebox{90}{biased sampling \quad\quad random sampling}
    \end{minipage}\hspace{-.2cm}
    \begin{minipage}{.24\linewidth}
        \centering
        $N_\text{ED} = 10$
        \includegraphics[width=\linewidth,trim=20 0 15 0]{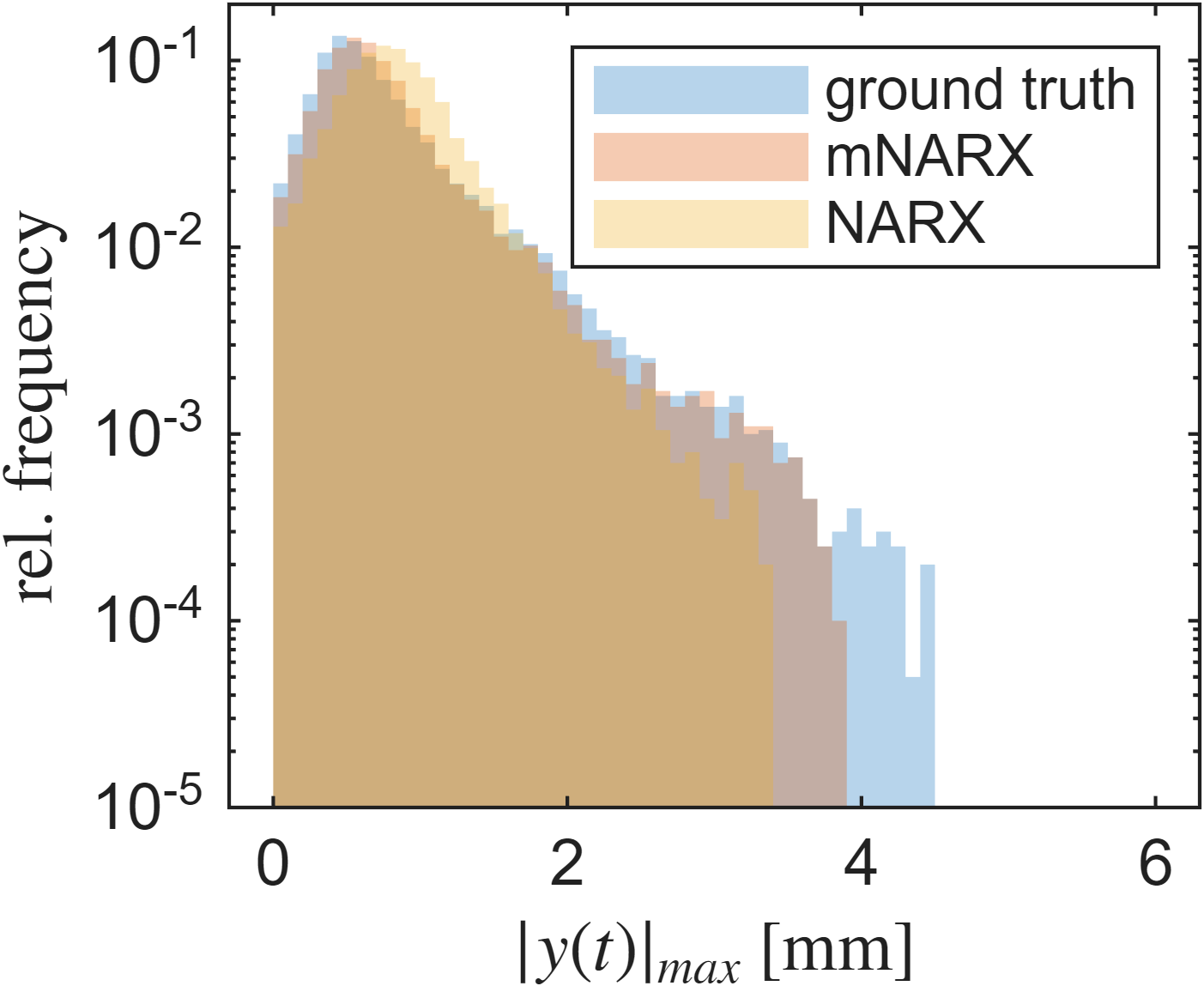}
        \includegraphics[width=\linewidth,trim=20 0 15 0]{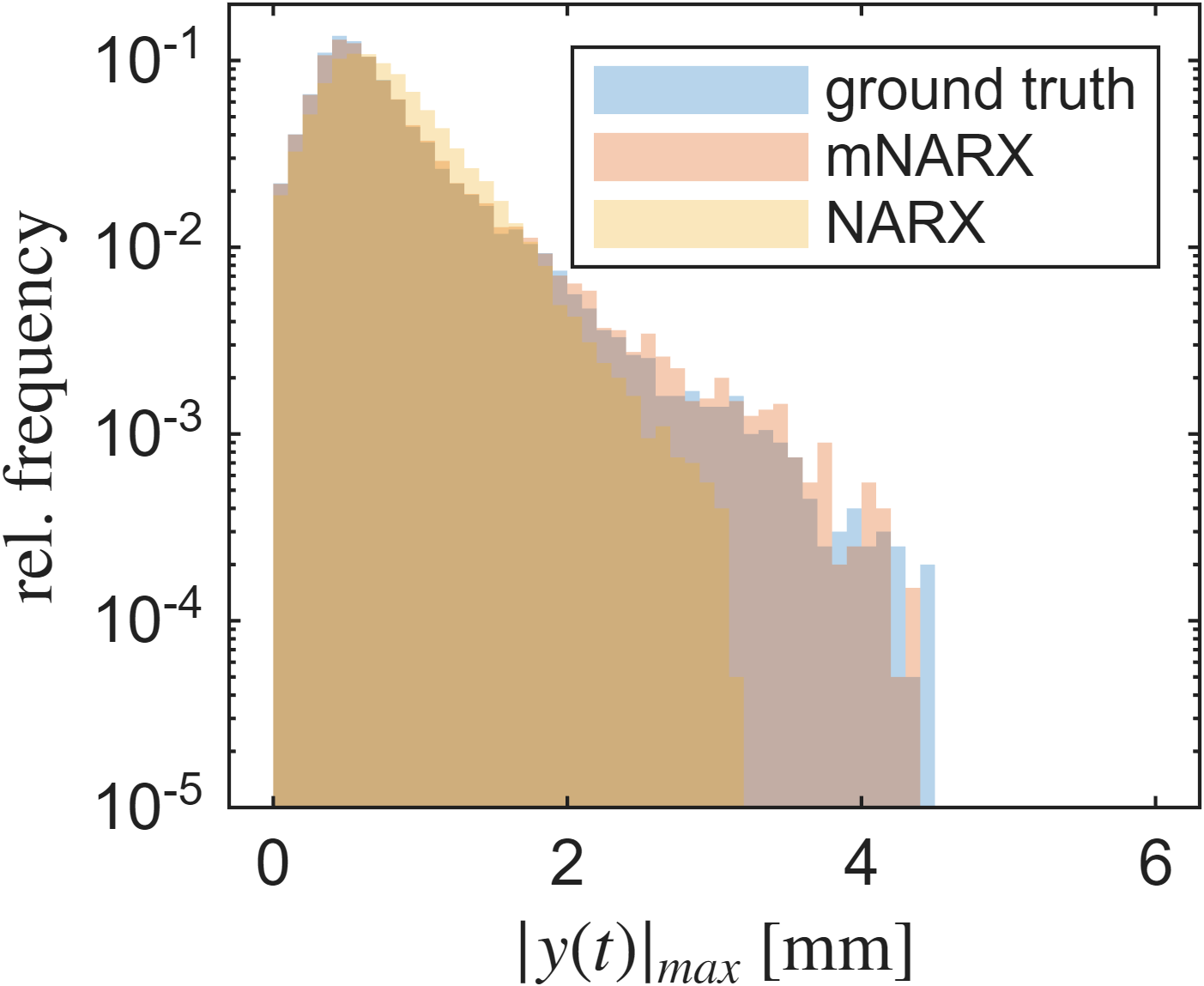}
    \end{minipage}    
    \hfill
    \begin{minipage}{.24\linewidth}
        \centering
        $N_\text{ED} = 50$
        \includegraphics[width=\linewidth,trim=20 0 15 0]{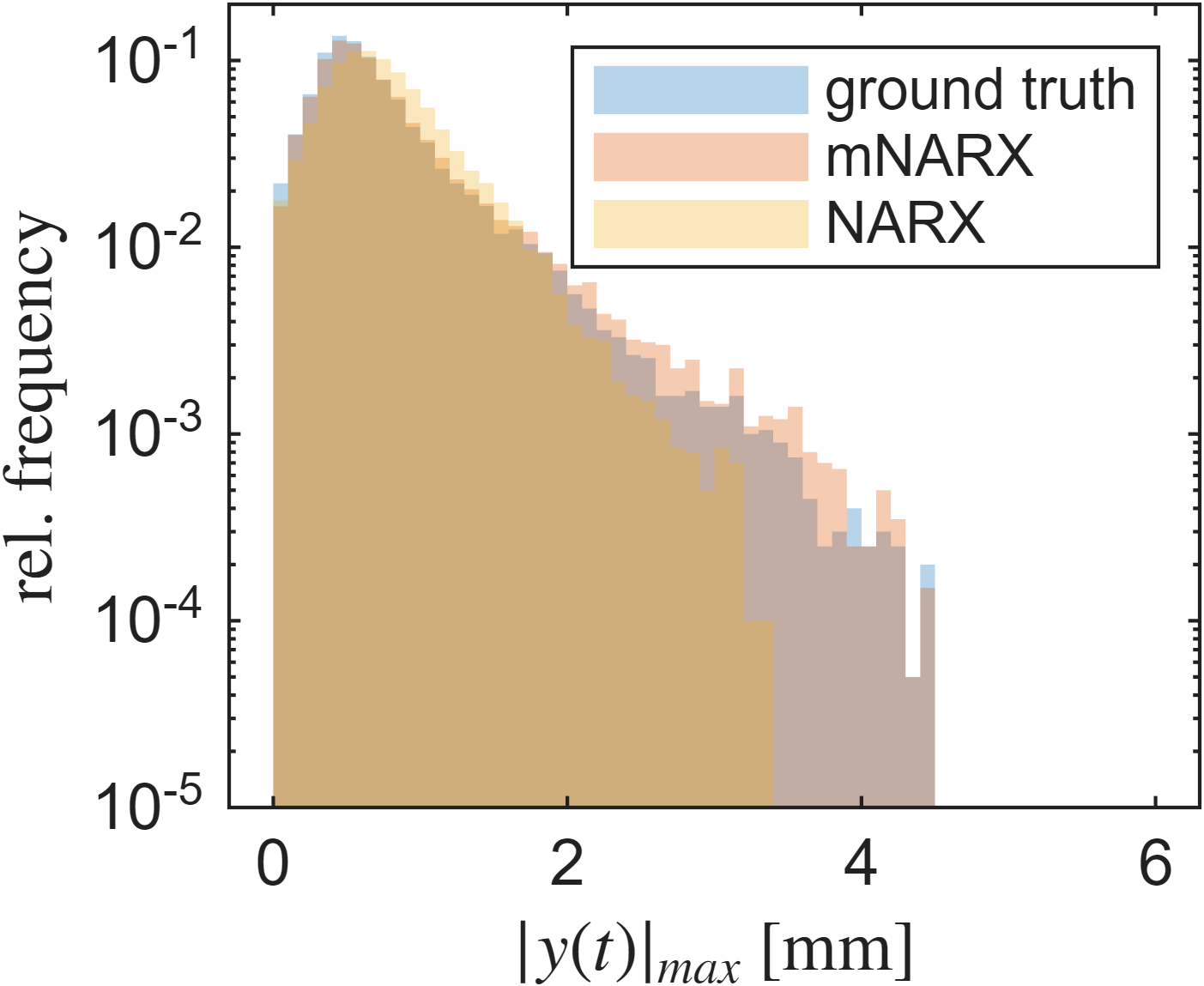}
        \includegraphics[width=\linewidth,trim=20 0 15 0]{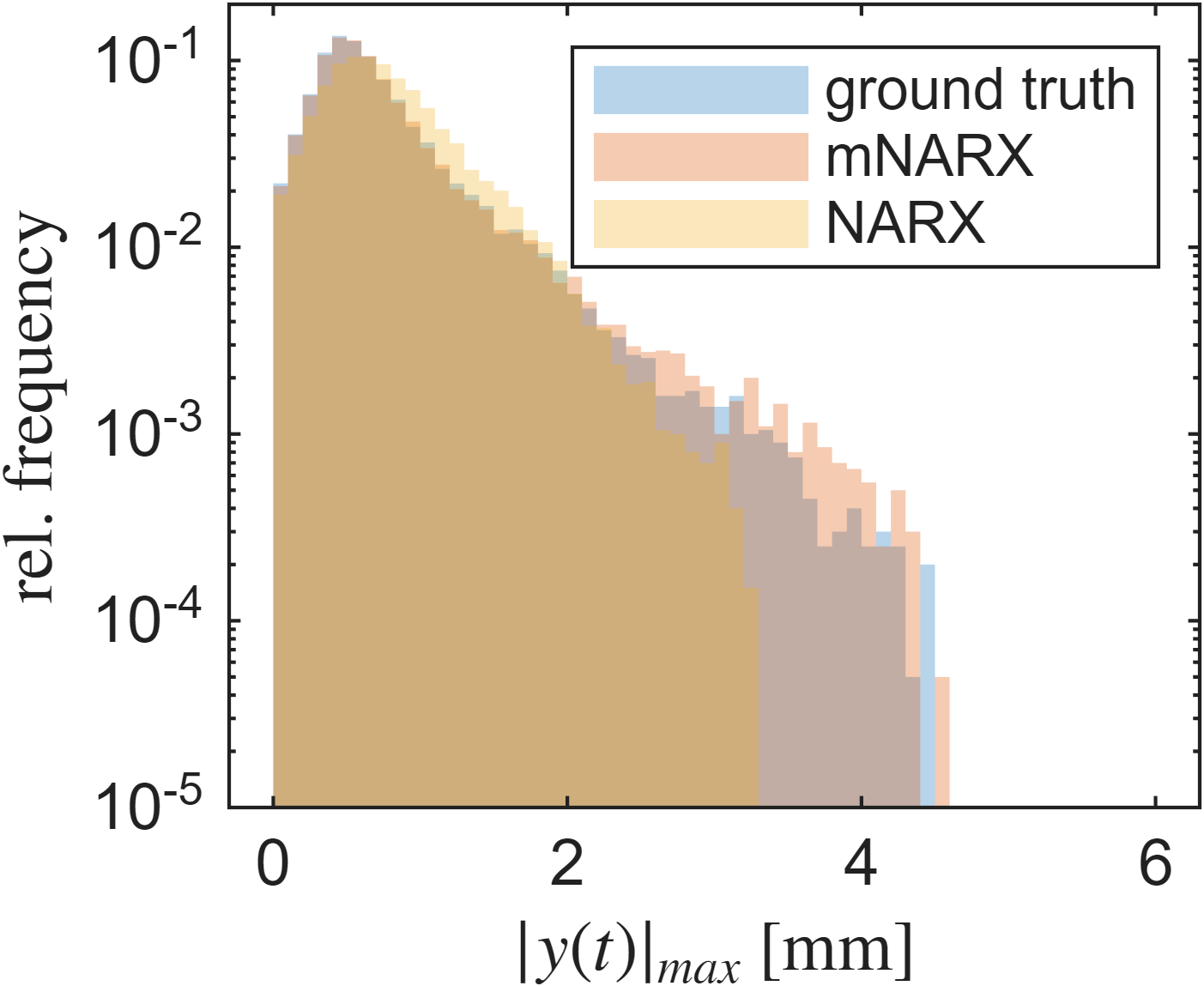}
    \end{minipage}    
    \hfill
    \begin{minipage}{.24\linewidth}
        \centering
        $N_\text{ED} = 100$
        \includegraphics[width=\linewidth,trim=20 0 15 0]{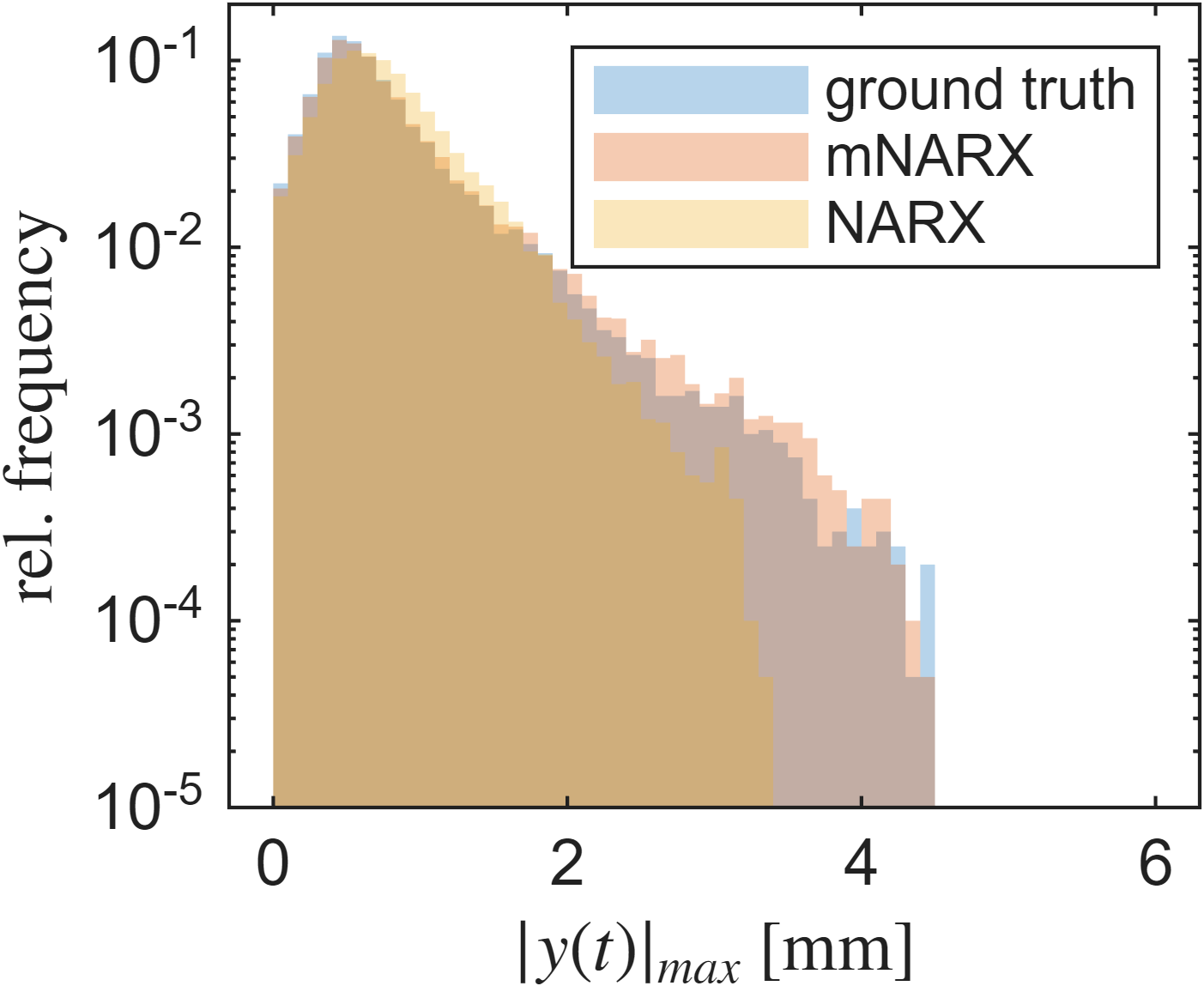}
        \includegraphics[width=\linewidth,trim=20 0 15 0]{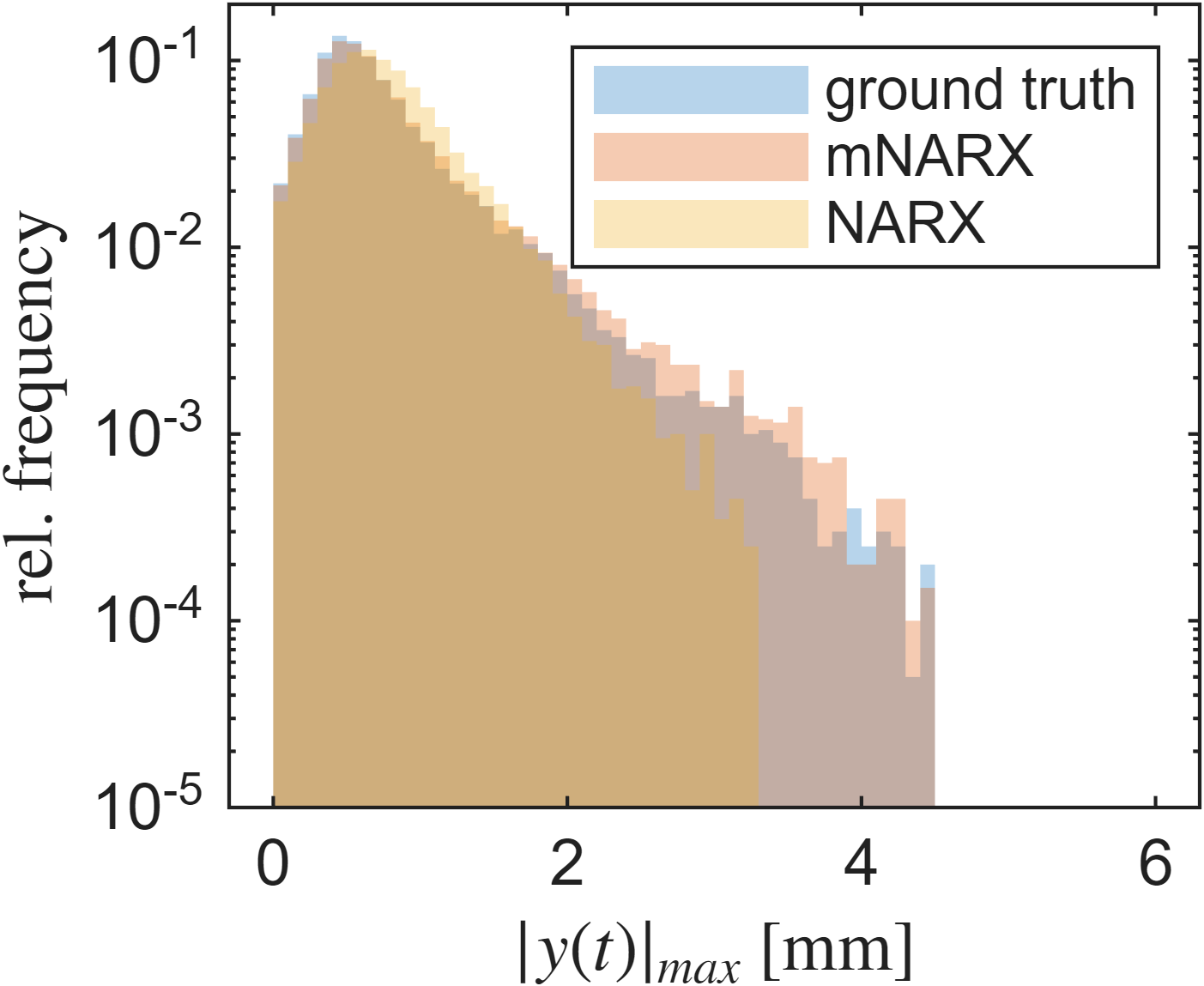}
    \end{minipage}    
    \caption{Empirical PDFs of the QoI $y_2(t)$ for different experimental design sizes $N_\text{ED}$, comparing unbiased and biased sampling strategies.}
    \label{fig:mNARX quarter car CDF}
\end{figure}

\begin{figure}[tb]
    \centering
      \begin{minipage}{.1\linewidth}
        \centering
        \rotatebox{90}{biased sampling \quad\quad random sampling}
    \end{minipage}   
    \begin{minipage}{.25\linewidth}
        \centering
        $N_\text{ED} = 10$
        \includegraphics[width=\linewidth,trim=20 0 15 0]{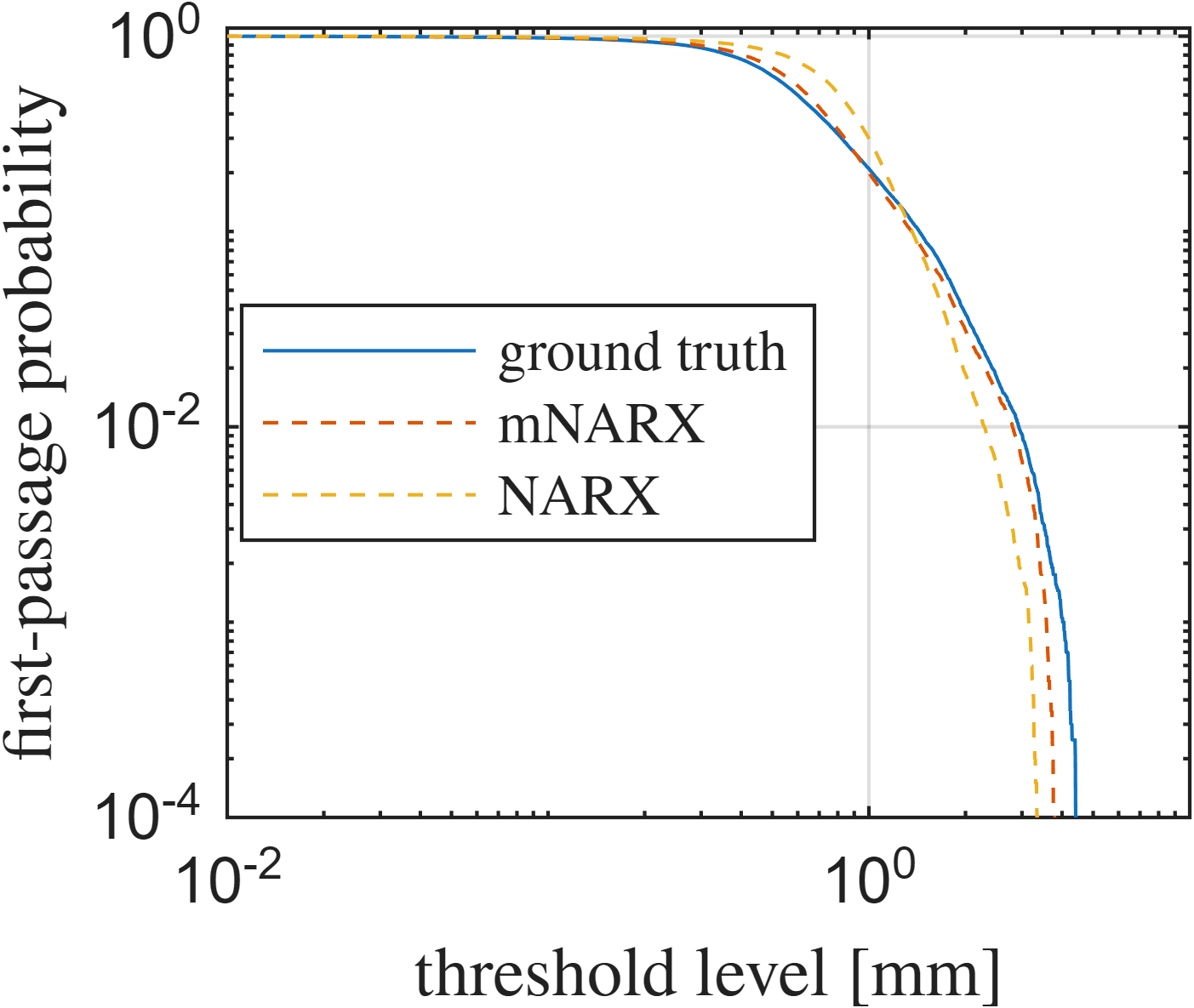}
        \includegraphics[width=\linewidth,trim=20 0 15 0]{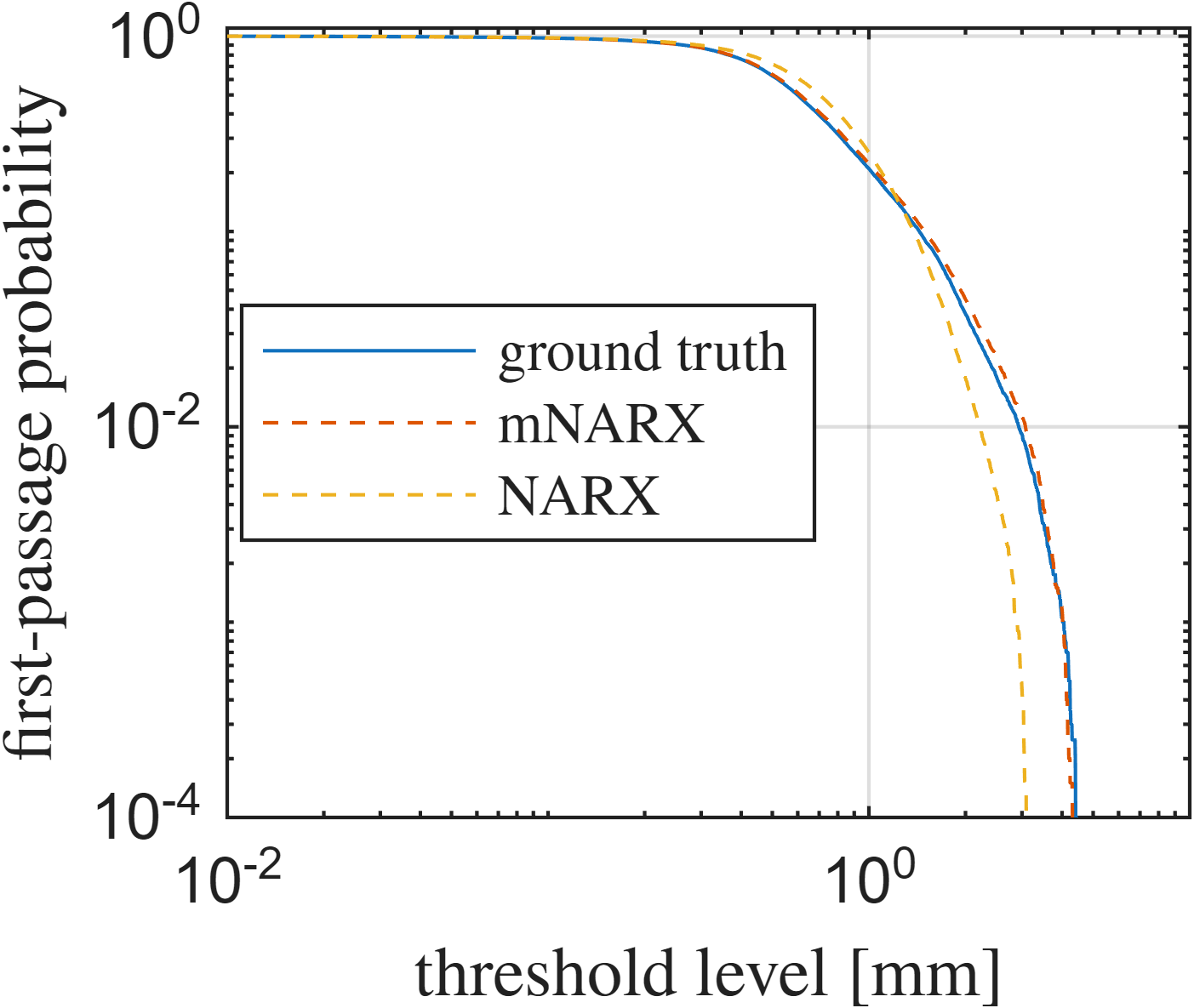}
    \end{minipage}
    \hfill
    \begin{minipage}{.25\linewidth}
        \centering
        $N_\text{ED} = 50$
        \includegraphics[width=\linewidth,trim=20 0 15 0]{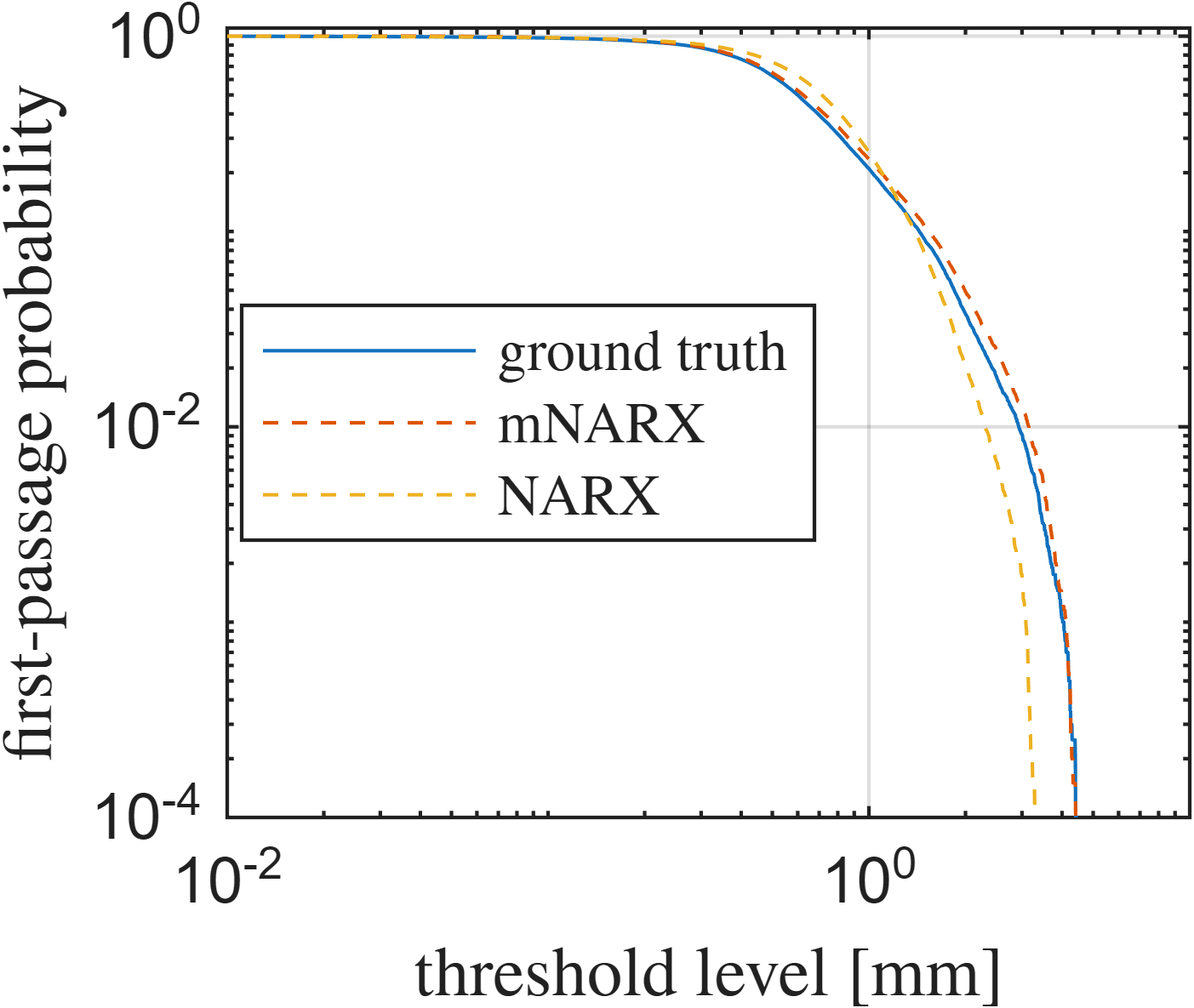}
        \includegraphics[width=\linewidth,trim=20 0 15 0]{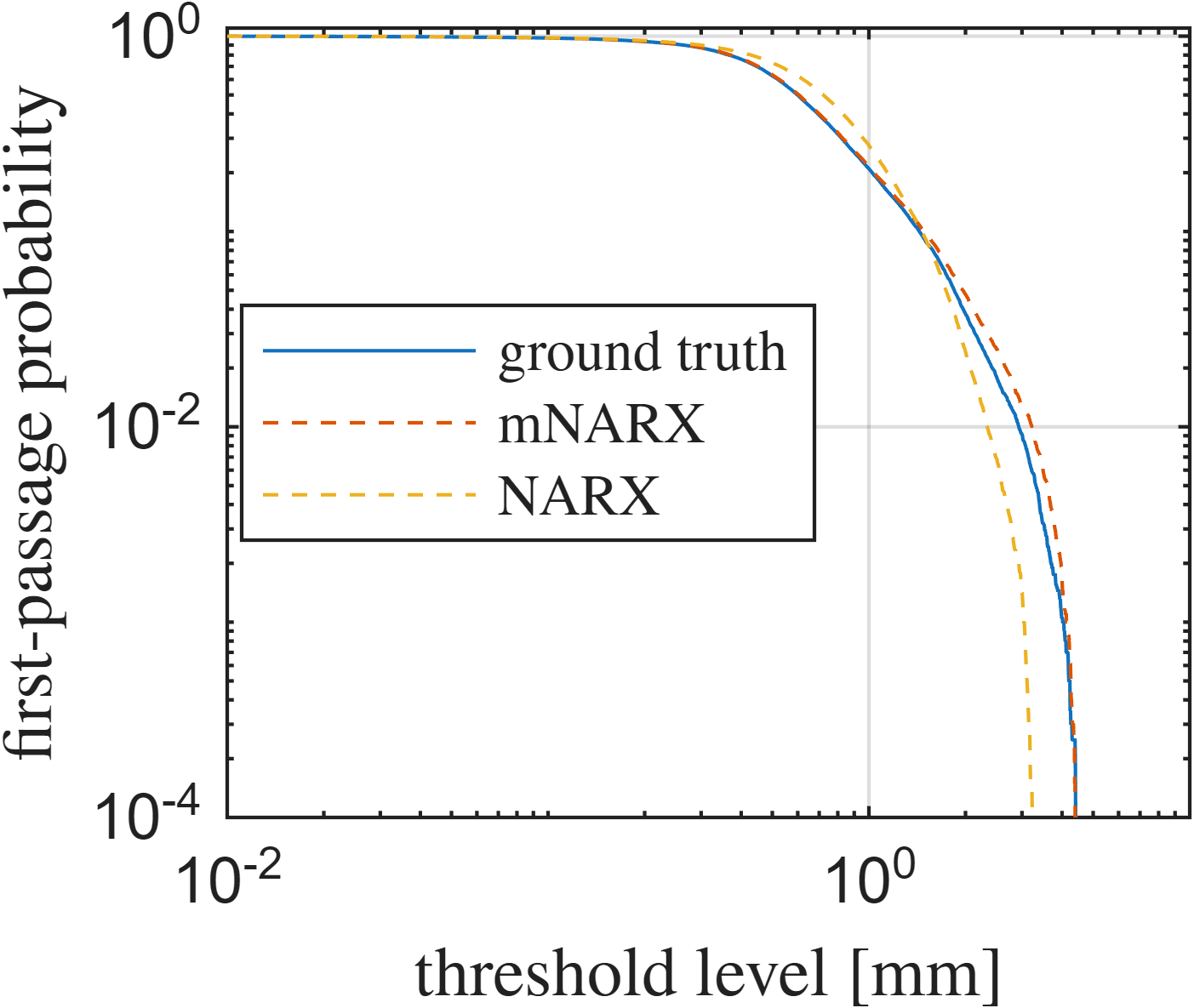}
    \end{minipage}    
    \hfill
    \begin{minipage}{.25\linewidth}
        \centering
        $N_\text{ED} = 100$
        \includegraphics[width=\linewidth,trim=20 0 15 0]{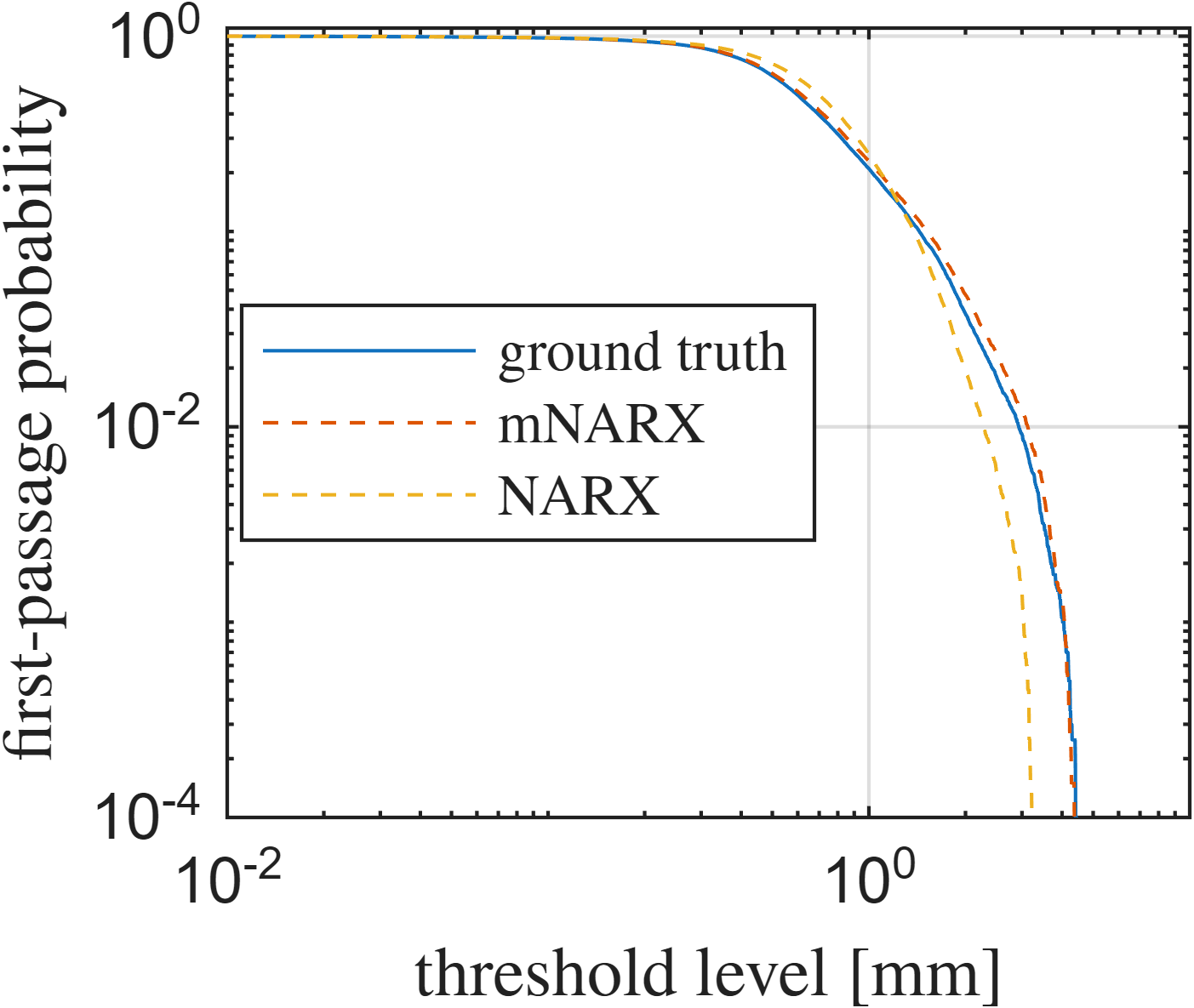}
        \includegraphics[width=\linewidth,trim=20 0 15 0]{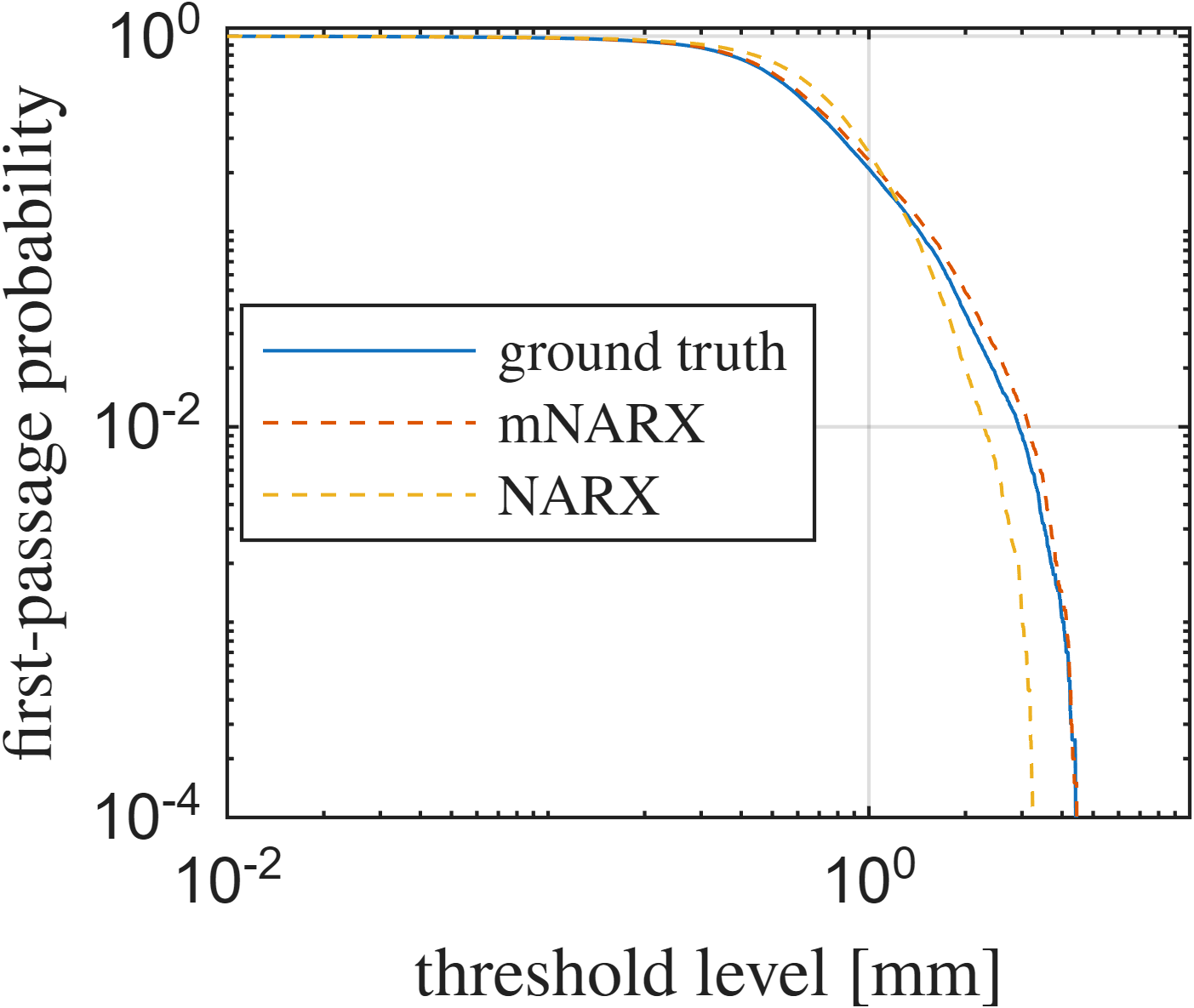}
    \end{minipage}   
    \caption{First-passage probability curves as a function of the threshold level for the quarter car model, obtained with both unbiased (top row) and biased (bottom row) sampling strategies for different values of $N_\text{ED}$, compared to the reference (continuous line). While in both cases the accuracy improves with $N_\text{ED}$, the biased sampling strategy provides a more accurate estimation of the first-passage time distribution, particularly in the tails, even for the smallest experimental design considered. When using biased sampling, the mNARX model is already accurate to within 5\% for a reference first-passage reliability index of $\beta = 3$ with as few as $N_\text{ED} = 10$ simulations in the experimental design.}
    \label{fig:mNARX quarter car FPT}
\end{figure}

Because the primary objective is to estimate small failure probabilities, we next evaluate the accuracy of the mNARX and NARX surrogates in estimating the response distribution tails.
Figure~\ref{fig:mNARX quarter car CDF} compares the empirical cumulative distribution functions (CDFs) of the QoI $y_{2}^{\max} = \max|y_2(t)|$, obtained with both unbiased (first row) and biased (second row) sampling strategies for different values of $N_\text{ED}$.
The results show that the both sampling strategies can achieve remarkable accuracy with as few as $O(100)$ samples even in a reliability analysis setting. 
Additionally, biased sampling strategy significantly outperforms standard random subsampling, especially in the tails of the distribution for small experimental design sizes, a critical feature for reliability analysis.
The classical NARX model performs poorly regardless of the sampling strategy and experimental design size, indicating a lack of model expressivity.
Finally, we assess the accuracy of the mNARX and NARX surrogates in estimating the first-passage time distributions, which are crucial for time-variant reliability analysis.
Figure~\ref{fig:mNARX quarter car FPT} compares the empirical first-passage probability for different thresholds (in abscissa), obtained with both unbiased (top row) and biased (bottom row) sampling strategies for different values of $N_\text{ED}$.
The results confirm that the biased sampling strategy provides a more accurate estimation of the first-passage time distribution, particularly in the tails, even for the smallest experimental design considered ($N=10$ training samples).
Also, this figure makes it clear that the NARX model is already insufficiently accurate even for relatively high first-passage probabilities of $0.1$.
 \clearpage
\section{Method II: Functional feature-based NARX ($\mathcal{F}$-NARX)}
\label{sec:FNARX}

While the mNARX framework effectively handles high-dimensional inputs through manifold construction, it still relies on the selection of discrete time lags to define the autoregressive terms. This dependence on discrete lags poses specific challenges:
\begin{itemize}
    \item \textbf{Lag Selection Difficulty:} Determining the optimal set of lags for systems with long memory or high sampling rates is nontrivial and often computationally expensive.
    \item \textbf{Discrete-Time Limitations:} Standard NARX models can suffer from numerical instability when the time step $\Delta t$ is small, as adjacent time steps are highly correlated, leading to ill-conditioned regression matrices.
\end{itemize}

To overcome these issues, we introduce here the recently developed \emph{functional feature-based NARX} ($\mathcal{F}$-NARX) model \citep{schar2025FNARX}. 
This second approach shifts the perspective from discrete time points to continuous time windows, exploiting the functional nature of the system's trajectory.

\subsection{From Discrete Lags to Functional Features}

In the classical NARX formulation, the prediction at time $t$ depends on a vector of past values at specific discrete lags, as in Eq.~\eqref{eq:NARX}. In contrast, $\mathcal{F}$-NARX postulates that the system response depends on the trajectory of the input and output over a recent memory window $\eta(t, T) = [t-T, t]$.

\begin{figure}[tb]
    \centering
    \includegraphics[width=0.7\textwidth]{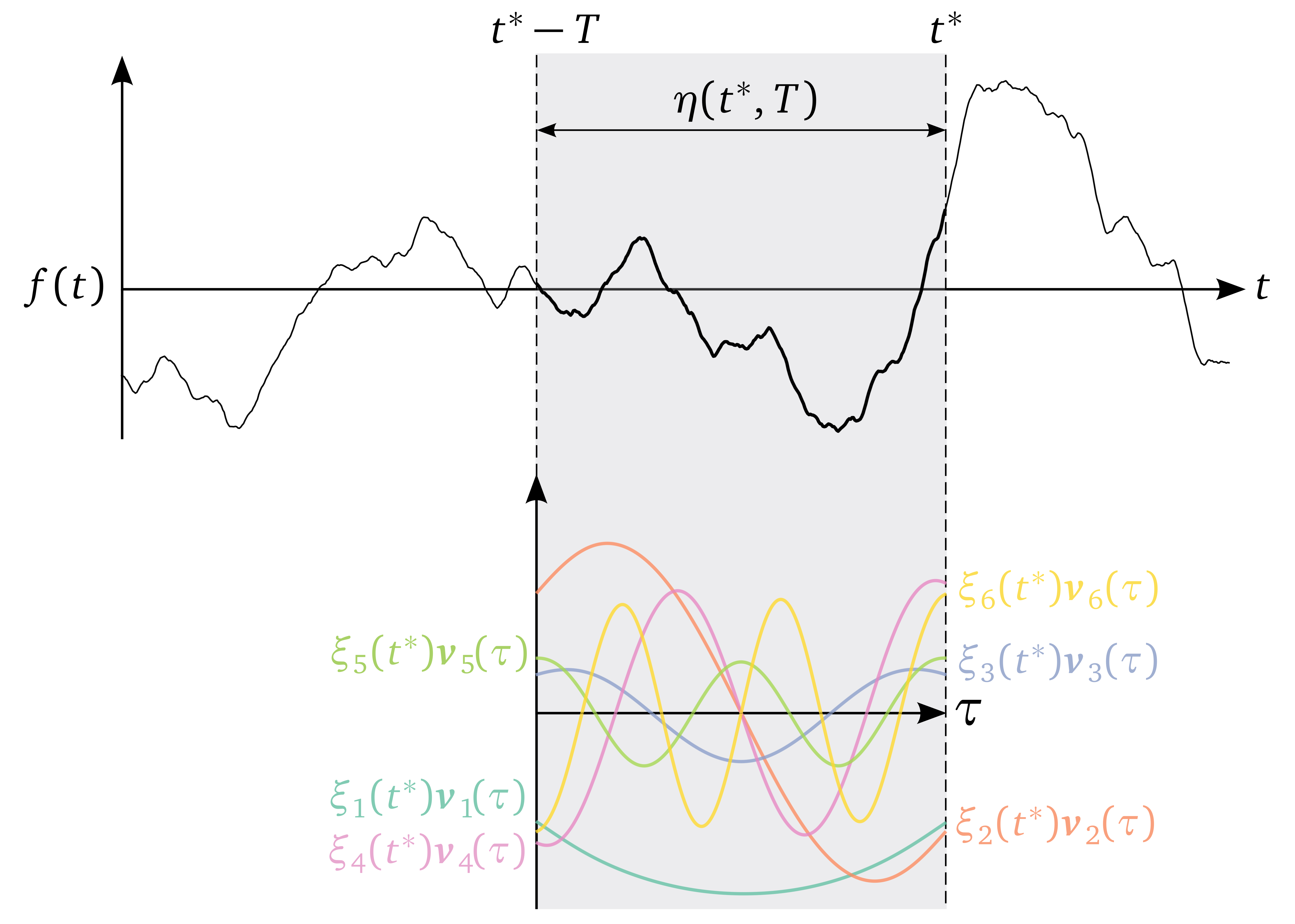}
    \caption{Graphical representation of the feature extraction process at the core of the $\mathcal{F}$-NARX method. An input or output trajectory over the memory window $[t-T, t]$ is transformed into feature vectors $\vXi_{\vx}(t)$ and $\vXi_y(t)$, which are then used in the regression model to predict $y(t)$ (adapted from \cite{schar2025FNARX}).}
    \label{fig:FNARX_features}
\end{figure}

Instead of selecting individual points within this window, $\mathcal{F}$-NARX extracts \emph{features} that summarize the trajectory's behavior. The surrogate model takes the form:
\begin{equation}
    y(t) \approx \widehat{\cm} \left( \vXi_{\vx}(t), \vXi_y(t) \right),
\end{equation}
where $\vXi_{\vx}(t)$ and $\vXi_y(t)$ are feature vectors representing the history of the input and output, respectively, over the memory window.
This is represented graphically in Figure~\ref{fig:FNARX_features}.

\subsection{Feature Extraction via PCA}

A common and robust method for extracting features from time series is principal component analysis (PCA), also known as proper orthogonal decomposition (POD) in time series analysis. 
In the context of $\mathcal{F}$-NARX, it can be directly applied to the time-windowed trajectories.

Consider  the lagged matrix $\ve{\Phi}_i$ of the $i^\text{th}$ variable $\in \{\vx_1 \cdots \vx_M, \,\vy\}$, based on Eq.~\eqref{eq:Phi}:
\begin{equation}
    \ve{\Phi_i} = \begin{bmatrix}\phi_i(t_{\text{min}}) \\ \phi_i(t_{\text{min}}+\delta t) \\ \vdots \\ \phi_i(t_{N}) \end{bmatrix} 
    , 
\end{equation} 
with  $\phi_i(t) = \left[x_i(t), x_i(t-\delta t), \dots, x_i(t-n_x\delta t) \right]$ for the exogeonous inputs, and $\phi_{M+1}(t) = \left[y(t-\delta t), \dots, y(t-n_y\delta t) \right]$ for the autoregressive component.\\
The PCA-based feature extraction process involves the following steps:
\begin{enumerate}
    \item Standardize $\ve{\Phi}_i$ by subtracting the column-wise mean $\ve\mu_i$ and dividing by the corresponding column-wise standard deviation $\ve\sigma_i$:
    \begin{equation*}
        \ve{Z}_i = \frac{\ve{\Phi}_i - \ve\mu_i}{\ve\sigma_i}
    \end{equation*}
    \item Compute the temporal covariance matrix across the available data:
    \begin{equation*}
        \ve{C}_i = \frac{1}{\tilde{N} - 1} \ve{Z}_i^\top \ve{Z}_i.
    \end{equation*}
    \item Compute its eigenvalue decomposition, to obtain a set of eigenvalues $\lambda_{ij}$ and eigenvectors $\ve{v}_{ij}$:
    \begin{equation*}
        \ve{C}_i \ve{v}_{ij} = \lambda_{ij} \ve{v}_{ij}
    \end{equation*}
    \item After sorting them in decreasing order, retain the first $\tilde n_i$ eigenvectors:
    \begin{equation*}
        \ve{V}_i = \{ \ve{v}_{i1}, \ve{v}_{i2}, \ldots, \ve{v}_{i\tilde{n}_i} \},
    \end{equation*}
    and related eigenvalues $\ve{\Lambda}_i = \text{diag}(\lambda_{i1}, \lambda_{i2}, \ldots, \lambda_{i\tilde{n}_i})$.

    The number of modes is chosen with a threshold (typically $0.99$) on the cumulative explained variance ratio $\nu_i$;
    \begin{equation*}\label{eq:explained_variance}
        \nu_i = \frac{\sum_{k=1}^{\tilde{n}_i} \lambda_{ik}}{\sum_{\ell=1}^{n_i} \lambda_{i\ell}}.
    \end{equation*}
\item Finally, project any signal for a new time window on the selected eigenvectors $\vXi_i$ by projection: 
    \begin{equation*}\label{eq:pca_mapping}
    \vXi_i = \mathcal{K}_i^\text{PCA}(\ve{\Phi}_i) = \ve{\Phi}_i \ve{V}_i .
\end{equation*}
\end{enumerate}


The same procedure is then applied to the all input and response time series $\boldsymbol x(t)$ and $\boldsymbol y(t)$ to obtain $\vXi_x(t)$ and $\vXi_y(t)$.

This transformation decouples the \emph{length} of the system memory (determined by $T$) from the \emph{complexity} of the regression model (determined by the number of retained eigenvectors $|\vXi_x(t)|$ and $|\vXi_y(t)|$). 
Even for systems with relatively long memories, the trajectory is often smooth and can be accurately represented by a small number of features (e.g., $\tilde n_i \approx 3-5$), significantly reducing the dimensionality of the regression problem compared to standard lag-based NARX.
In addition, because the features depend on the smoothness of the trajectory, rather than its discretization, the $\mathcal{F}$-NARX model is more robust to oversampling and numerical instabilities (see detailed discussion in \cite{schar2025FNARX}).

\subsection{Application: The Bouc-Wen Oscillator}\label{sec:Bouc Wen}

\begin{figure}[tb]
    \centering
    \begin{minipage}{.49\linewidth}
            \centering
            \includegraphics[width=.8\linewidth]{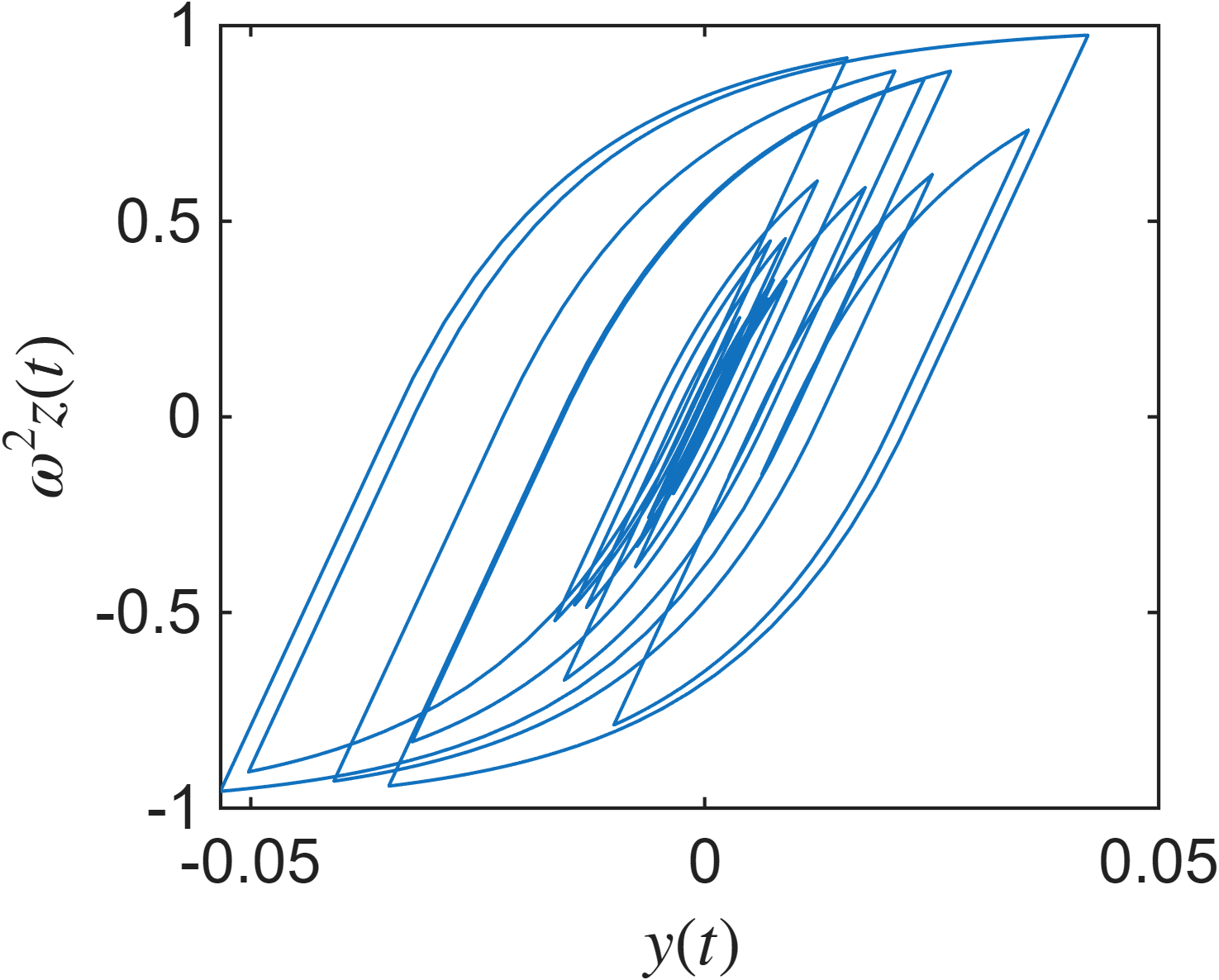}
    \end{minipage}
    \begin{minipage}{.49\linewidth}
            \centering
            \begin{tabular}{lcc}
                \hline
                Parameter & Unit & Value \\ \hline
                $\zeta$   & -                        & 0.02                      \\
                $\omega$  & rad/s                    & 10                        \\
                $\rho$    & -                        & 0.2                       \\
                $\gamma$  & -                        & 0.5                       \\
                $\alpha$  & 1/m                      & 25                        \\
                $\beta$   & -                        & 25                        \\
                $n$       & -                        & 1                         \\ \hline
            \end{tabular}
    \end{minipage}
    \caption{Left: graphical illustration of the hysteretic behavior of the Bouc-Wen oscillator described in Section~\ref{sec:Bouc Wen}. Right: model parameters used in the numerical example.}
    \label{fig:bouc wen}
\end{figure}

To demonstrate the performance of the $\mathcal{F}$-NARX method combined with mNARX for 
reliability analysis, we consider the Bouc-Wen oscillator in Figure~\ref{fig:bouc wen}. 
It is a hysteretic system widely used in structural dynamics to model nonlinear material behavior. 
The corresponding governing equations read:
\begin{equation}
    \begin{cases}
        \ddot{y}(t) + 2\zeta\omega \dot{y}(t) + \omega^2 (\rho y(t) + (1-\rho) z(t)) = -x(t) \\
        \dot{z}(t) = \gamma \dot{y}(t) - \alpha |\dot{y}(t)| |z(t)|^{n-1} z(t) - \beta \dot{y}(t) |z(t)|^n
    \end{cases}
    \label{eq:boucwen}
\end{equation}
where $z(t)$ is the hysteretic variable that tracks the system's history dependence. The parameters of the Bouc-Wen model are given in Figure~\ref{fig:bouc wen}.
While the system is considered deterministic, it is subjected to a stochastic ground motion excitation $x(t)$ modeled by a filtered white noise process through the Rezaeian - Der Kiureghian stochastic ground motion model \citep{rezaeian2010_core}. 
Its parameters are estimated from component 090 of the Northridge earthquake recorded at the LA 00 station \citep{chu_2016_core} and are provided in Table~\ref{tab:ground_motion_parameters}. 

\begin{table}[tb]
    \caption{Parameters of the Rezaeian - Der Kiureghian stochastic ground-motion model in Section~\ref{sec:Bouc Wen} (from \cite{mai_2016}). }
    \label{tab:ground_motion_parameters}
    \centering
    \begin{tabular}{lcc}
        \hline
        Parameter & Unit & Value \\ \hline
        $I_a$ & $s\cdot g$ & $0.109$ \\
        $D_{5-95}$ & s & $7.96$ \\
        $t_{\text{mid}}$ & s & $7.78$ \\
        $\omega_\text{mid}$ & Hz & $4.66 \times 2\pi$ \\
        $\omega'$ & Hz & $-0.09 \times 2\pi$ \\
        $\zeta_\text{f}$ & - & $0.24$ \\ \hline
    \end{tabular}
\end{table}

This example is challenging both because of its nonlinearity, but especially because it is an hysteretic system that keeps memory of its past trajectory through the variable $z(t)$. This behavior is difficult to capture with autoregressive models, because the system response at time $t$ depends not only on the recent history of the input and output, but also on the entire past trajectory through $z(t)$. 
This makes it an ideal test case for the $\mathcal{F}$-NARX method, which can leverage the functional features to capture the long-memory effects without requiring an excessively large number of discrete lags, as well as mNARX, which can use the hysteretic variable $z(t)$ as a component of its input manifold.

\subsection{$\mathcal{F}$-NARX Configuration}
Following the previous example application on the quarter-car model, we construct an {$\mathcal{F}$-NARX} surrogate to approximate the oscillator displacement $y(t)$ based on the ground acceleration input $x(t)$. 
We consider here the same two training strategies for the experimental design: plain and biased sampling (see Section~\ref{sec:mNARX}).
As in mNARX, we include the hysteretic variable $z(t)$ as an additional output to be modeled, which is then used as an auxiliary input when predicting $y(t)$.
The configuration of the $\mathcal{F}$-NARX model is summarized in Table~\ref{tab:FNARX Bouc-Wen configuration}.

\begin{table}[tb]
    \centering
    \caption{$\mathcal{F}$-NARX configuration parameters for the Bouc-Wen model.}
    \label{tab:FNARX Bouc-Wen configuration}\small
  \begin{tabular}{lcccc}
        \toprule
        \rowcolor[HTML]{FFFFFF} 
        \multicolumn{1}{l}{\cellcolor[HTML]{FFFFFF}}  & \multicolumn{2}{c}{\textbf{Biased sampling}} & \multicolumn{2}{c}{\textbf{Random sampling}} \\ \midrule
        $\vy(t)$                              & $z(t)$               & $y(t)$                            & $z(t)$          & $y(t)$          \\
        \rowcolor[HTML]{FFFFFF} 
        $\boldsymbol x$                      & $\ddot{x}(t), \dot{x}(t), x(t)$ & $\ddot{x}(t), \dot{x}(t), x(t), \hat{z}(t)$ & $\ddot{x}(t), \dot{x}(t), x(t)$ & $\ddot{x}(t), \dot{x}(t), x(t), \hat{z}(t)$ \\
        T                   & 1 sec. / 50 lags     & 1 sec. / 50 lags                 & 1 sec. / 50 lags & 1 sec. / 50 lags \\
        $\nu_i$                   & 90 \%                 & 90 \%                             & 90 \%            & 90 \%            \\
        \textbf{Degree}            & 2                     & 3                                 & 2                & 3                \\
        \textbf{Interaction}                    & 2                     & 2                                 & 2                & 2                \\ 
        \bottomrule
        \end{tabular}
    
\end{table}

\subsection{$\mathcal{F}$-NARX - Reliability analysis}
\begin{figure}[tb]\centering
    \includegraphics[width=0.4\linewidth]{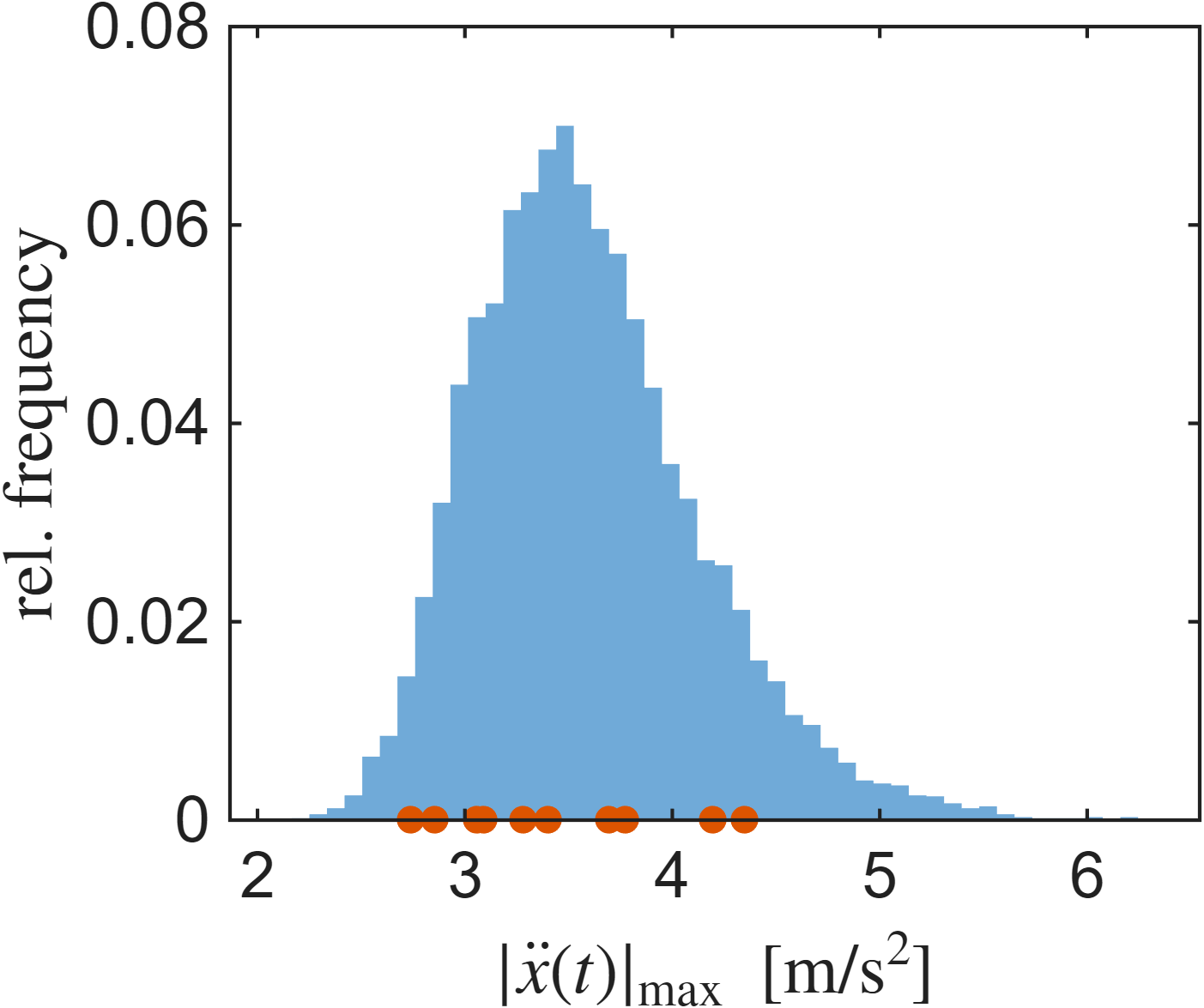}   
    \includegraphics[width=0.4\linewidth]{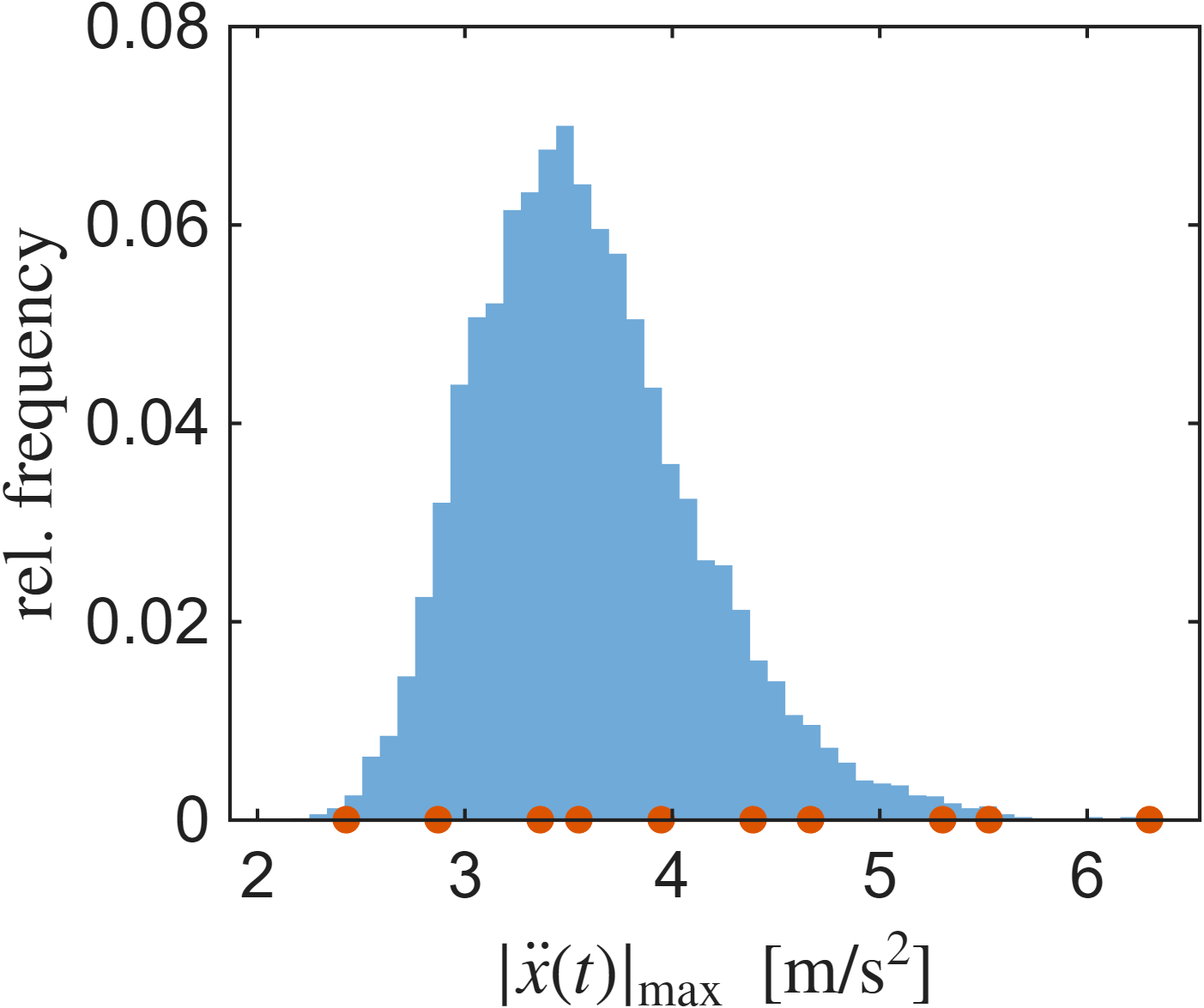}
    \caption{Graphical depiction of the biased sampling strategy in Section~\ref{sec:mNARX biased ED}. In both panels, the histogram shows the distribution of maximum absolute excitation amplitudes $|x(t)^{(i)}|_{\max}$ for a large candidate set of $10^4$ input realizations, while the red dots represent the selected amplitudes. Left: standard random subsampling; Right: the biased selection ensures a much wider coverage of extreme amplitudes.}
    \label{fig:FNARX biased ED}
\end{figure}

For this analysis, we consider a single, comparatively small experimental size $N = 50$, with the two different sampling strategies described in Section~\ref{sec:mNARX biased ED}.
We obtain the biased sample by selecting input realization that result in a uniformly distributed maximum stochastic acceleration $\max |\ddot{x}(t)|$ out of a large candidate set of $10^4$ realizations. 
The bias selection is represented graphically in Figure~\ref{fig:FNARX biased ED}.

\begin{figure}[tb]\centering
        \begin{minipage}{0.45\linewidth}
        \centering
        \includegraphics[width=\linewidth]{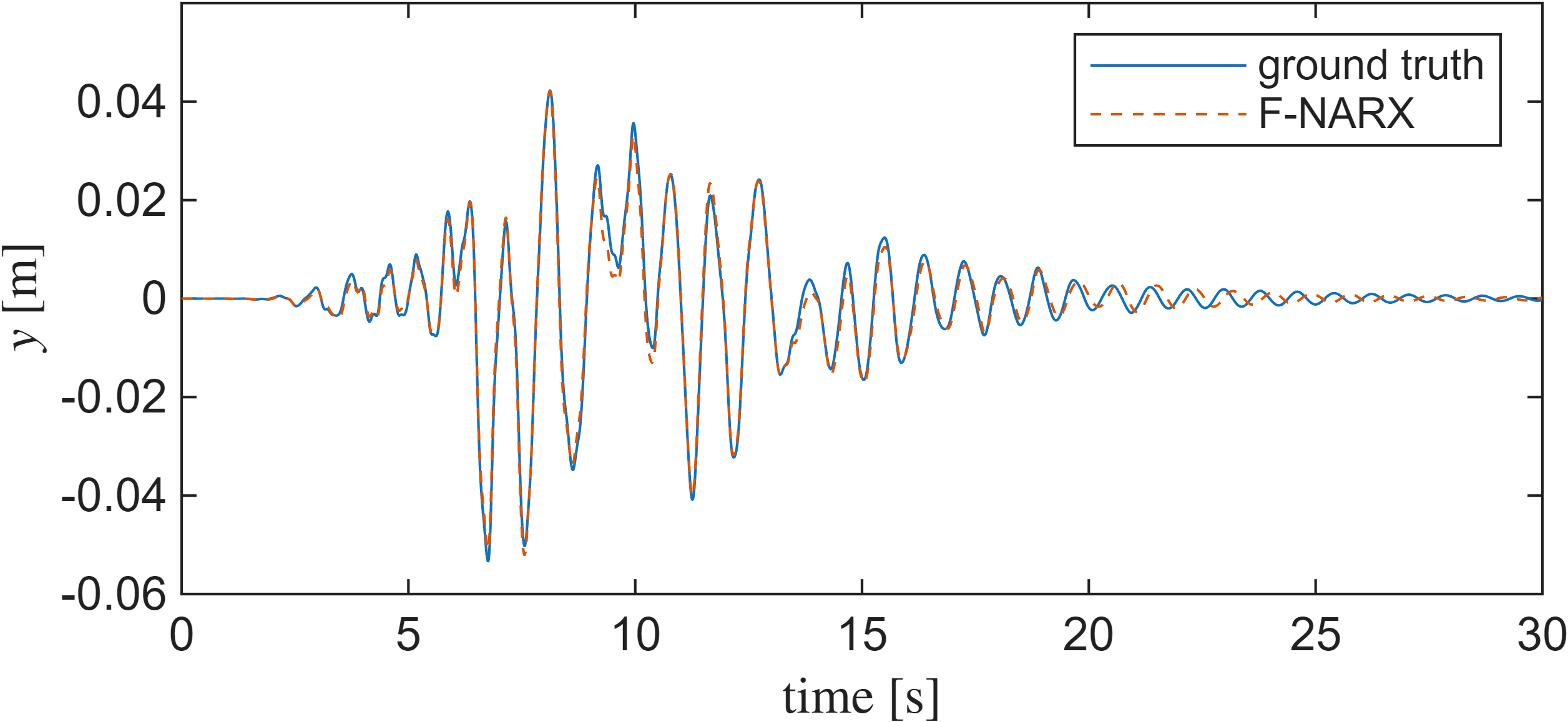}
        \includegraphics[width=\linewidth]{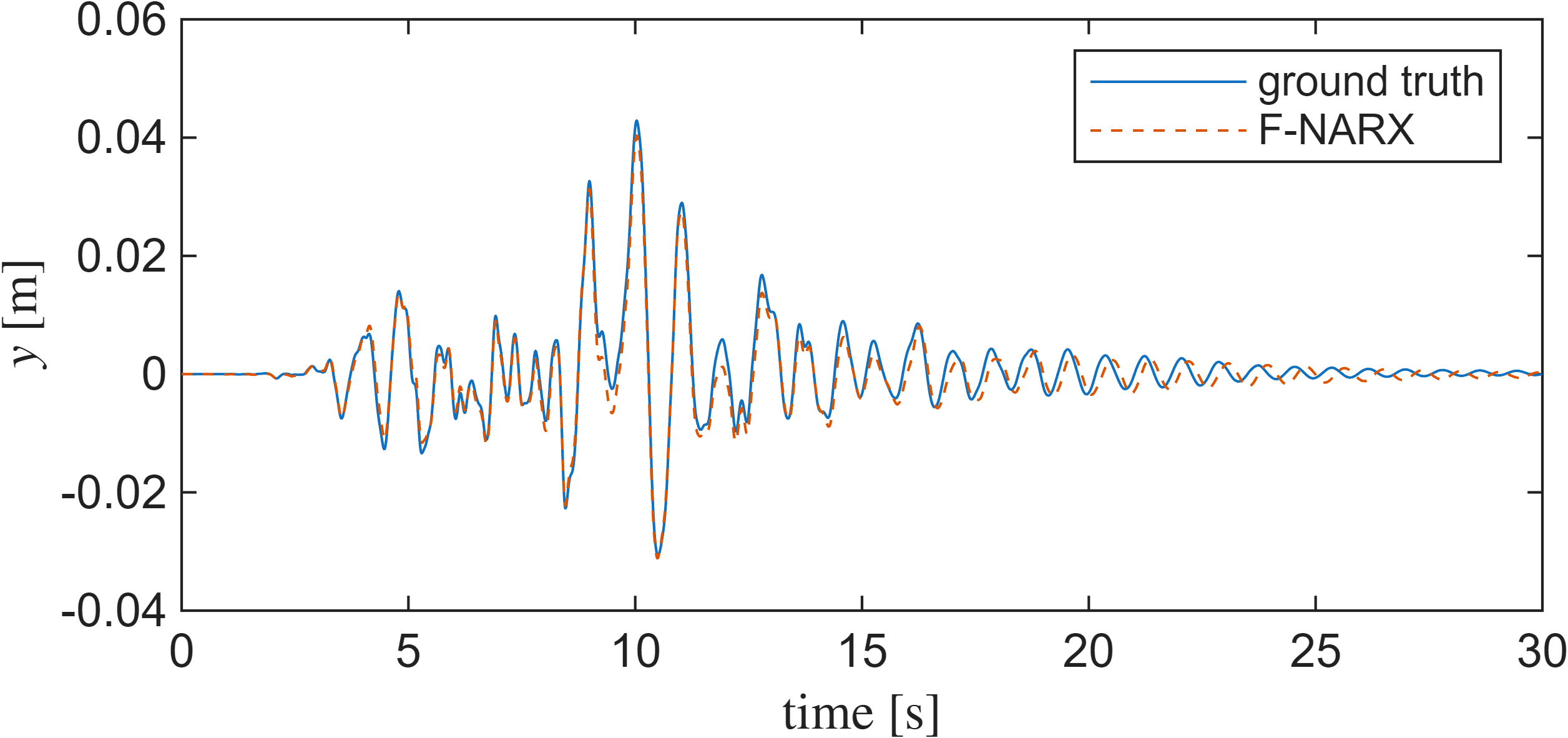}
    \end{minipage}
    \begin{minipage}{0.45\linewidth}
        \centering
        \includegraphics[width=\linewidth]{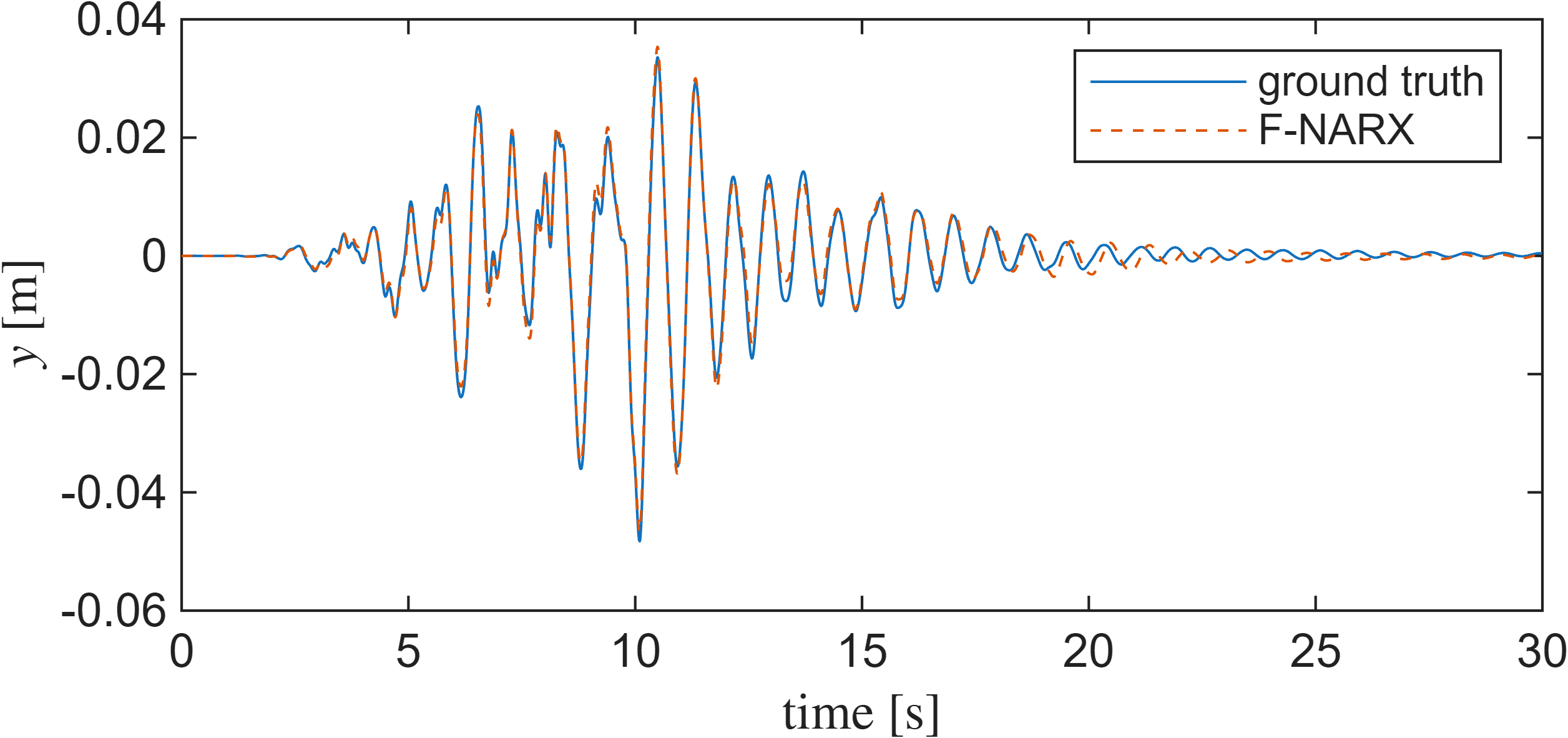}
        \includegraphics[width=\linewidth]{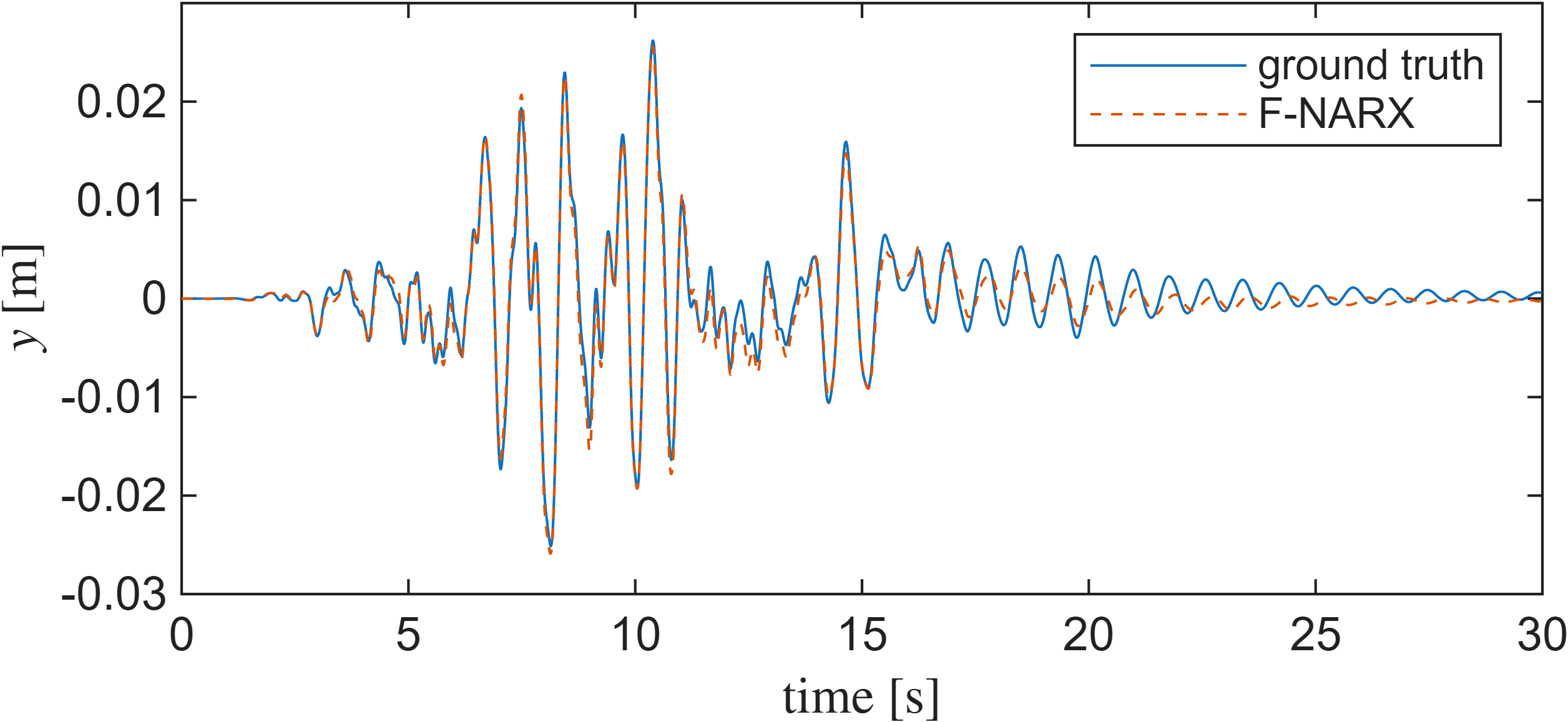}
    \end{minipage}
    \caption{Selected out-of-sample traces comparing the ground-truth Bouc-Wen response (solid blue lines) with the $\mathcal{F}$-NARX predictions (dashed orange lines) for different input realizations, using 50 biased samples. The $\mathcal{F}$-NARX model accurately captures the hysteretic behavior of the system across a range of excitation levels.}
    \label{fig:bouc wen traces}
\end{figure}

\begin{figure}[tb]
    \begin{minipage}{.49\linewidth}
        \centering
        Random sampling ($N_\text{ED}=50$)
        
        \includegraphics[width=0.85\linewidth]{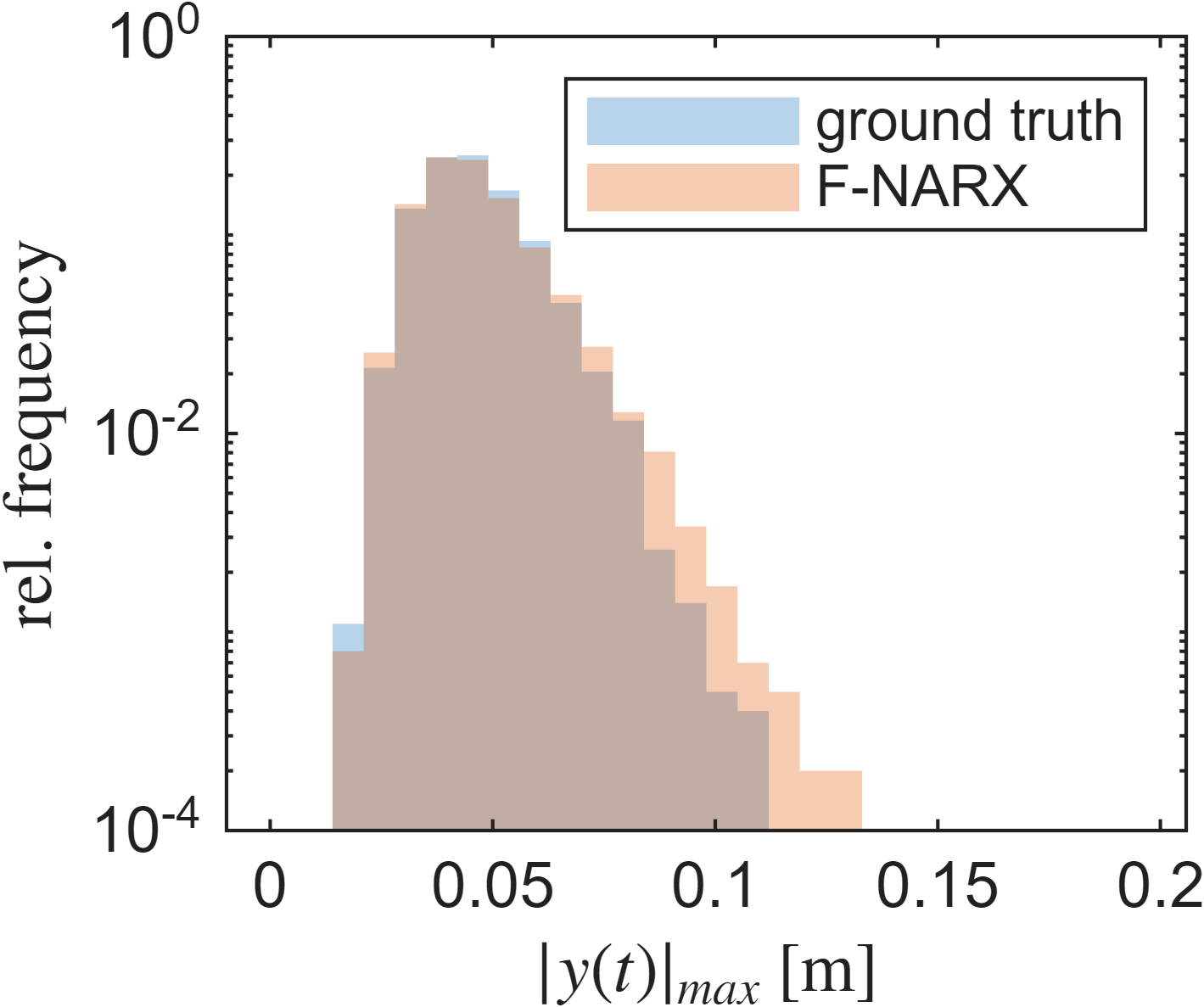}
    \end{minipage}
    \begin{minipage}{.49\linewidth}
        \centering
        Biased sampling ($N_\text{ED}=50$)
        
        \includegraphics[width=0.85\linewidth]{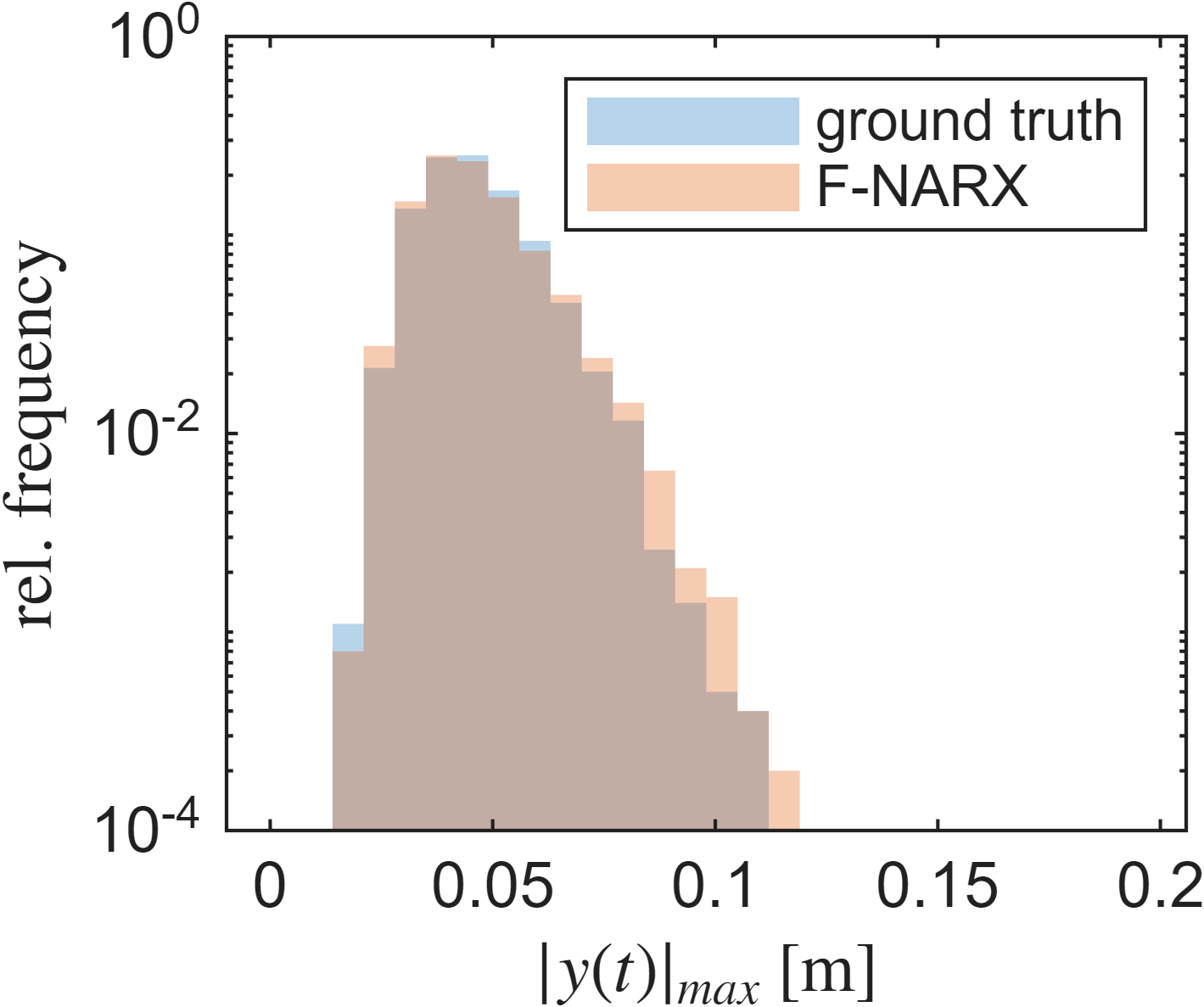}
    \end{minipage}    \\
    \begin{minipage}{.49\linewidth}
        \centering        
        \includegraphics[width=0.8\linewidth]{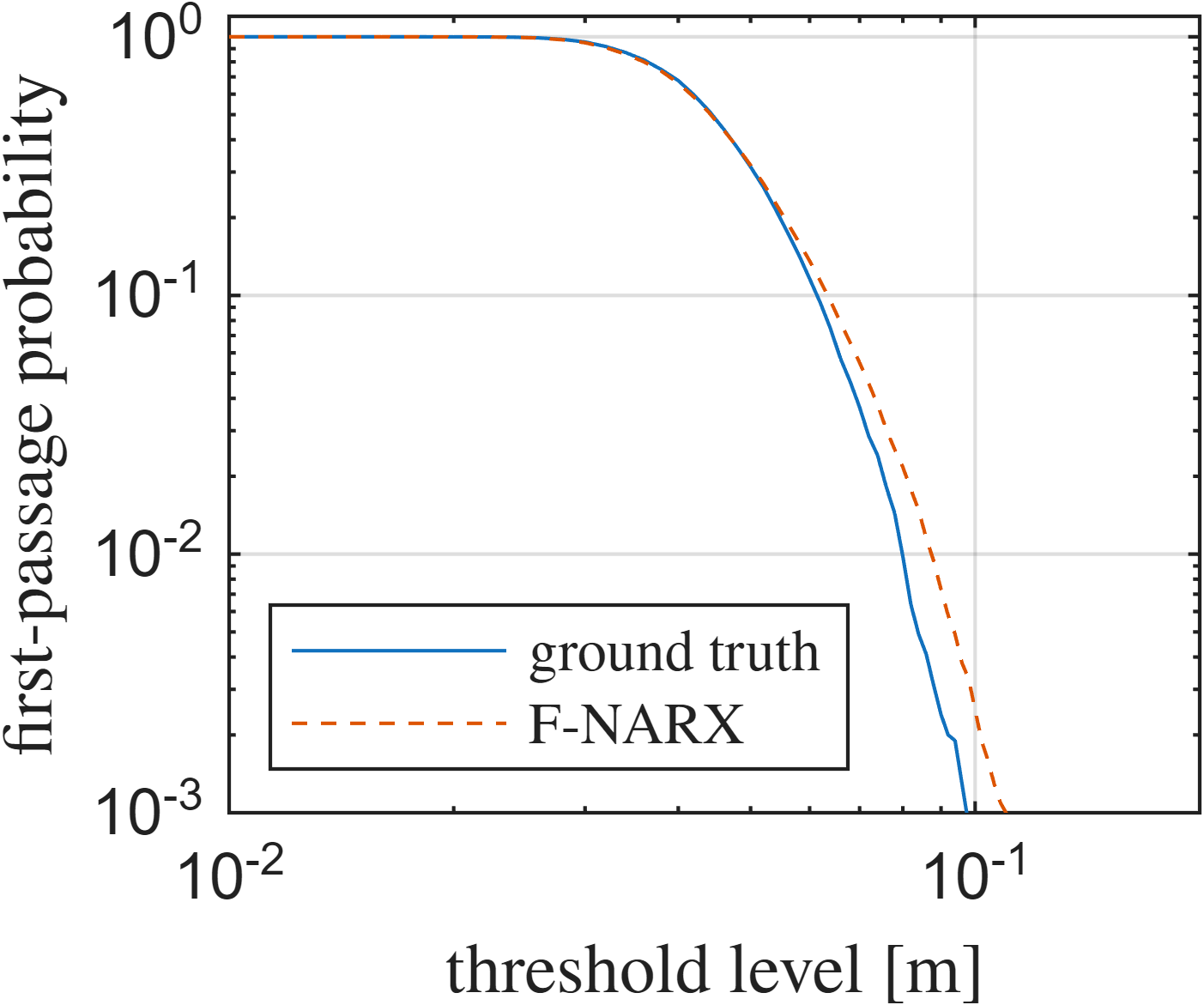}\\[0.2em]
    \end{minipage}
    \begin{minipage}{.49\linewidth}
        \centering  
        \includegraphics[width=0.8\linewidth]{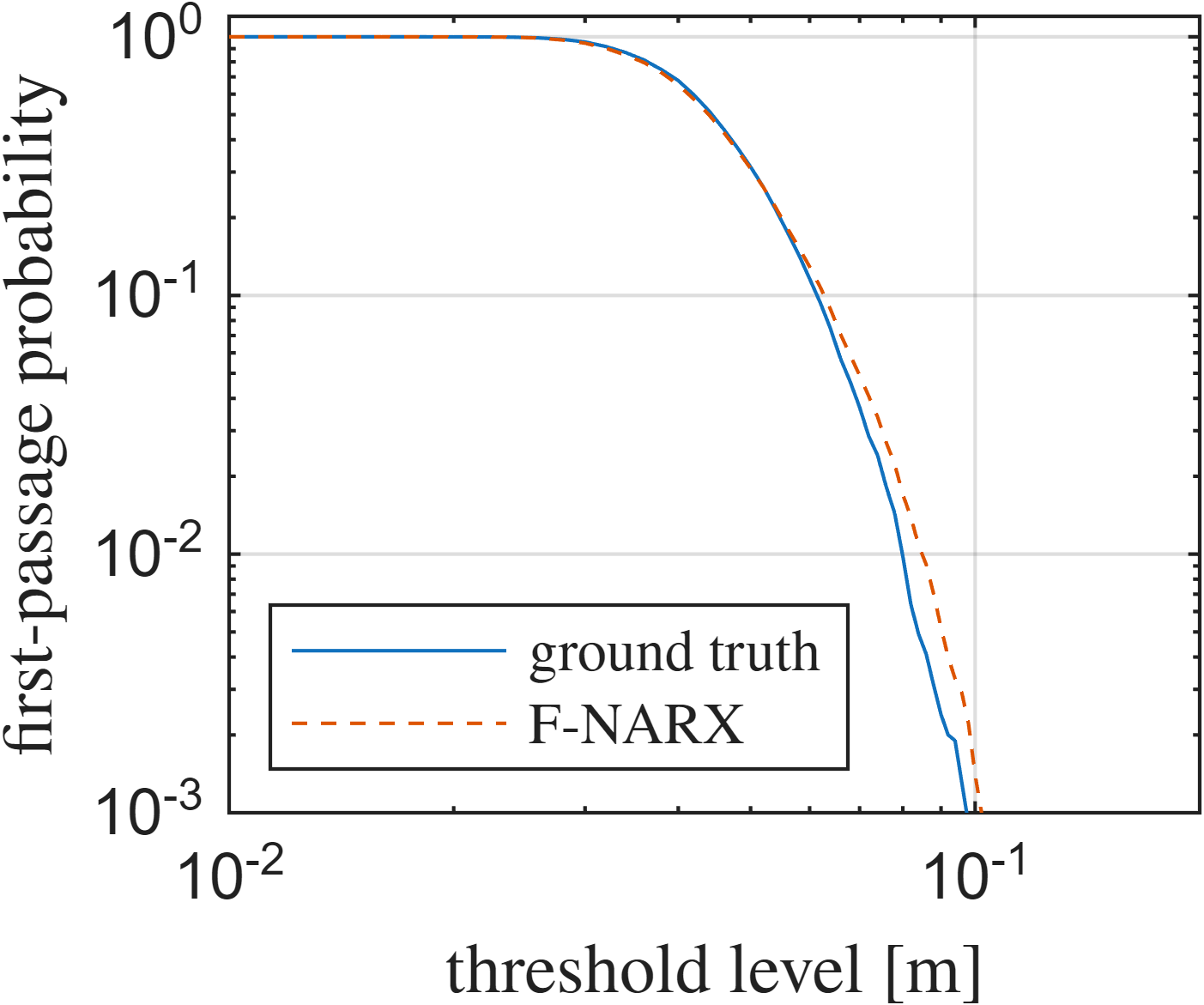}\\[0.2em]
    \end{minipage}    
    \caption{Response histograms and first-passage probability estimates for the Bouc-Wen oscillator using $\mathcal{F}$-NARX surrogates trained on $N_\text{ED}=50$ samples. Left: results using random sampling; Right: results using biased sampling. Top row: histograms of maximum displacement $\max |y(t)|$; Bottom row: estimated first-passage probability $P_f(t)$ as a function of the admissible threshold. Although not perfectly accurate, the $\mathcal{F}$-NARX model with biased sampling better captures the tail behavior of the response distribution with respect to the standard sampling, leading to more reliable first-passage estimates. The average error on a first passage reliability index $\beta_{f,c} = 3$ is in the 10-15\% range, even with such a small experimental design.}
    \label{fig:FNARX results}
\end{figure}

The resulting calibrated $\mathcal{F}$-NARX model is used to estimate the full time history of the oscillator displacement, as well as the corresponding first-passage probability. 
A set of traces comparing out-of-sample full model responses with the predictions of the $\mathcal{F}$-NARX model trained on the biased experimental design is shown in Figure~\ref{fig:bouc wen traces}. The traces demonstrate that the $\mathcal{F}$-NARX model accurately captures the hysteretic behavior of the system across a range of excitation levels, even when trained on a relatively small number of full model evaluations.
The corresponding histograms of the maximum displacement response $\max |y(t)|$ and the estimated first-passage probability $P_f(t)$ are shown in Figure~\ref{fig:FNARX results}.

The results show good agreement with the reference simulations, even for rare failure events (${P_f \approx 10^{-3}}$), where the biased $\mathcal{F}$-NARX model consistently estimates the first-passage reliability index $\beta_{f,c} \equiv -\Phi^{-1}(P_f)$ to within 15\% of the reference value $\beta_{f,c} = 3$, despite the relatively small experimental design. The $\mathcal{F}$-NARX model successfully captures the hysteretic nature of the response without explicitly modeling the internal variable $z(t)$, proving that the functional features contain sufficient information to reconstruct the system's path-dependent state. Furthermore, the use of biased sampling for training (as introduced in Section~\ref{sec:mNARX}) is again shown to be crucial for accurate tail estimation.

\section{Discussion and conclusions}
\label{sec:Conclusion}

Time-variant reliability analysis imposes a severe computational burden on the design of modern engineering systems. The necessity to estimate first-passage probabilities for rare failure events requires evaluating complex dynamical models over thousands of stochastic excitation scenarios. This chapter has presented a methodology to overcome this bottleneck by replacing the expensive transient solver with specialized time-dependent surrogate models.

We first discussed the limitations of classical time-invariant surrogates and standard NARX models, particularly in the case of high-dimensional inputs and highly nonlinear dynamics. 
These difficulties are further compounded by the complexities of choosing appropriate discrete time lags, as well as surrogate modeling parameters. 
To address these challenges, we adopted two recently developed surrogate modeling frameworks: manifold NARX (mNARX) and functional feature-based NARX ($\mathcal{F}$-NARX). Both methods reformulate the NARX approach to better capture the temporal dependencies and nonlinearities inherent in dynamical systems. More specifically: 
mNARX constructs the surrogate on a manifold augmented with auxiliary intermediate variables, effectively ``unfolding'' the nonlinearity of the system. This method is particularly powerful when physical insight allows for the identification of key state variables (e.g., velocity or nonlinear internal forces) that simplify the mapping from excitation to response. 
$\mathcal{F}$-NARX shifts instead from discrete time lags to continuous time-window features, thus decoupling the memory length of the system from the complexity of the regression model. This makes it an ideal candidate for ``black-box'' systems exhibiting long memory or high sampling rates, where traditional lag selection is inefficient. 
A second important advantage of $\mathcal{F}$-NARX is how simple it is to parameterize with respect to standard NARX: the user only needs to specify the window length and the number of features, without having to worry about lag selection or stability issues. A common choice of window length is the characteristic time scale of the system, which can be estimated from the autocorrelation function of the response or from physical considerations. The number of features can be determined  by retaining a fixed percentage of the variance in the principal component analysis (PCA) step, typically in the order of $90-99\%$.

These results could be further enhanced in purely data-driven scenarios by adopting the recently proposed mNARX+ framework \citep{schar2026mnarx+}, which combines the strengths of both mNARX and $\mathcal{F}$-NARX by performing automatic feature extraction on the manifold-augmented space. This hybrid approach allows for the identification of informative features even when no physical insight is available, while still leveraging auxiliary variables to simplify the system dynamics.

While both mNARX and $\mathcal{F}$-NARX have shown promising results in the benchmark problems considered, it is important to note that their performance is highly dependent on the choice of experimental design. In particular, the use of biased sampling strategies that focus on the tail behavior of the response distribution is crucial for accurate estimation of first-passage probabilities. 
Preliminary research on the use of adaptive sampling techniques that iteratively refine the surrogate model based on the estimated failure probability have shown how $\mathcal{F}$-NARX can be effectively integrated into an active learning framework, further reducing the number of simulations required to achieve a given level of accuracy in the tail estimation \citep{song2026F2NARX}.

In conclusion, the proposed time-dependent surrogates bridge the gap between rigorous reliability theory and the practical constraints of computational dynamics, offering a path toward efficient design and safety assessment of complex, dynamical systems.

{\bf Acknowlegments ---} This work is part of the \emph{HIghly advanced Probabilistic design and Enhanced
  Reliability methods for high-value, cost- efficient offshore WIND} (HIPERWIND) project and has received
funding from the European Union’s Horizon 2020 Research and Innovation Programme under Grant Agreement
No. 101006689.

\bibliography{TimeDependentReliabilitySurrogates}

\end{document}